\documentclass[prl,twocolumn,showpacs,preprintnumbers,amsmath,amssymb,superscriptaddress]{revtex4-1}

\usepackage{graphicx}
\usepackage{dcolumn}
\usepackage{bm}
\usepackage{epsfig}
\usepackage{amsmath}
\usepackage{amssymb}
\usepackage{color}
\usepackage{bbm}
\usepackage[caption=false]{subfig}
\usepackage{placeins}
\usepackage{soul}

\usepackage[colorlinks=true,citecolor=blue,linkcolor=blue,urlcolor=blue]{hyperref}
\usepackage{cancel}

\def\equationautorefname~#1\null{%
	Eq.~#1\null
}
\def\figureautorefname~#1\null{%
	Fig.~#1\null
}

\begin{document}

\title{Exceptional Point Superradiant Lasing with Ultranarrow Linewidth }
\author{Min Du}
\affiliation{School of Physics and Institute for Quantum Science and Engineering, Huazhong University of Science and Technology, and Wuhan Institute of Quantum Technology, Wuhan 430074, China}

\author{Qian Bin}\email{qianbin@scu.edu.cn}
\affiliation{College of Physics, Sichuan University, Chengdu 610065, China}

\author{Qing-Yang Qiu}
\affiliation{School of Physics and Institute for Quantum Science and Engineering, Huazhong University of Science and Technology, and Wuhan Institute of Quantum Technology, Wuhan 430074, China}

\author{Franco Nori}
\affiliation{Theoretical Quantum Physics Laboratory, Cluster for Pioneering Research, RIKEN, Wako-shi, Saitama 351-0198, Japan}
\affiliation{Quantum Information Physics Theory Research Team, Quantum Computing Center, RIKEN, Wakoshi, Saitama, 351-0198, Japan}
\affiliation{Physics Department, The University of Michigan, Ann Arbor, Michigan 48109-1040, USA}

\author{Xin-You L\"{u}}\email{xinyoulu@hust.edu.cn}
\affiliation{School of Physics and Institute for Quantum Science and Engineering, Huazhong University of Science and Technology, and Wuhan Institute of Quantum Technology, Wuhan 430074, China}
\date{\today}

\begin{abstract}
	Achieving superradiant lasing with an ultranarrow linewidth is crucial for enhancing atomic clock stability in quantum precision measurement. By employing the exceptional point (EP) property of the system, we demonstrate theoretically superradiant lasing with linewidths in the $\mu$Hz range, sustained at the high-power level. This is achieved by incoherently pumping optical lattice clock transitions with ultracold alkaline-earth strontium-87 atoms in the EP of a $\mathcal{PT}$-symmetric system. Physically, the atomic coherence reaches a maximum in the EP, significantly amplifying the superradiance effect and resulting in superradiant lasing with an ultranarrow linewidth. This linewidth is even three orders of magnitude smaller than that of superradiant lasing in the systems without EP. Our work extends the realm  of superradiant lasing by introducing the EP property, and offers promising applications for developing atomic clocks with exceptional stability and accuracy.
\end{abstract}

\maketitle
As one of the most precise measuring instruments available, atomic clocks have broad application prospects in fields such as gravitational wave detection\,\cite{S. Kolkowitz}, dark matter exploration\,\cite{A. Derevianko}, relativity verification\,\cite{P. Delva}, measurement of fundamental physical constants\,\cite{R. M. Godun}, information networks\,\cite{J. Levine}, and satellite navigation\,\cite{L. Maleki}. With the advent of optical frequency combs, which enable the  measurement of optical frequencies with high precision\,\cite{T. Udem, Cundiff2003Ye},  it has become feasible to construct atomic clocks based on optical domain atomic transitions\,\cite{Ludlow2015BY}. Optical atomic clocks are currently classified into two main types according to their working mode: passive and active optical clocks\,\cite{J. Chen, Zhang2023SM}.  Current passive optical clocks operate by locking the frequency of an external optical local oscillator (ultra-stable laser) to the atomic transition. Clock performance depends critically on laser quality, making it sensitive to cavity length thermal noise.  In contrast, the active optical clocks are based on an ultra-stable optical frequency signal generated directly from an atomic system via stimulated emission. The optical frequency is defined by the atomic transition and is robust to the cavity noise. A superradiance laser-based optical clock is a typical active optical clock, where atoms collectively emit light to generate an optical frequency signal. The frequency linewidth of this clock depends on Purcell-enhanced atomic emission rate and can be ultranarrow, leading to the enhanced clock stability\,\cite{D. Meiser,H. Liu, M. J. Holland, M. A. Norcia, J. G. Bohnet}. In recent years, the extensive experimental and theoretical efforts have been dedicated to further narrowing the linewidth of superradiant lasers\,\cite{Matthew A. Norcia, T. Laske, S. Dubey, J. R. K. Cline, Y. Zhang, Yu. Huihui}, a crucial advancement for improving the frequency stability of atomic clocks.
\par
Exceptional points (EPs) are singularities in non-Hermitian parity-time ($\mathcal{PT}$) symmetric systems, where complex eigenvalues and their corresponding eigenvectors coalesce\,\cite{Bender1998Boettcher, El-Ganainy2018MK}. Typically, these singularities occur in systems with carefully designed gain and loss distributions. Recently, it has been proposed that purely dissipative systems can exhibit the hidden $\mathcal{PT}$ symmetry through gauge transformation\,\cite{S. K. Ozdemir}. EPs and $\mathcal{PT}$-symmetric systems have been widely studied in various platforms, including optical\,\cite{Guo2009SD, Ruter2010MEG, Peng2014OR, Feng2014WM, Hodaei2014MH},  electronics\,\cite{Schindler2011LZ}, microwaves\,\cite{Bittner2012, M. Fan}, mechanics\,\cite{Bender2013BP, X. Y. Lu}, acoustics\,\cite{Jing2014OL, Fleury2015SA}, and atomic\,\cite{Hang2013HK, Zhang2016,P. Peng} systems. In the EP, a $\mathcal{PT}$-symmetric system can undergo an abrupt phase transition,  losing $\mathcal{PT}$ symmetry and entering the $\mathcal{PT}$ symmetry-breaking region. Because of this unique property, EP can induce the intriguing phenomena such as loss-induced coherence\,\cite{Peng2014OR}, chiral mode switch\,\cite{Doppler2016MB}, and significant  sensitivity enhancement\,\cite{Zhong-Peng Liu, Z. Li, Jan. Wiersig, H. Hodae, W. Chen}, and so on. A natural  question is whether EPs could influence steady superradiant lasing, particularly in reducing linewidth—a critical feature for atomic clocks. Moreover, the crossover of $\mathcal{PT}$ symmetry theory and superradiant lasing in cavity quantum electrodynamics remains largely unexplored, which may substantially advance the field of lasing engineering and quantum metrology.
\par
Here, we propose to achieve superradiant lasing with an ultranarrow linewidth in a $\mathcal{PT}$-symmetric system consisting of two coupled dissipative cavities, one of which traps thousands of atoms. Interestingly, the EP property of the system can influence the atomic coherence and Purcell enhanced atomic emission rate, which reduces the upper threshold of the collective radiation regime. In the EP, the atomic coherence reaches its peak, significantly amplifying the superradiance effect and lastly resulting in the superradiant lasing with an ultranarrow linewidth---three orders of magnitude narrower than that of superradiant lasing in the systems without EP. The linewidth of the superradiant lasing can reach the scale of $\mu$Hz, which is also much smaller than the individual atomic decay rates. This work provides a new approach for enhancing the stability and accuracy of atomic clocks, and  such advancements have potential applications in the international time standards\,\cite{G. Panfilo}, the national time services\,\cite{S.C.Li}, the satellite navigation systems\,\cite{B. Jaduszliwer}, and the research on general relativity\,\cite{J. C. Hafele}.
\par
\begin{figure}
	\centering
	\includegraphics[width=7.8cm]{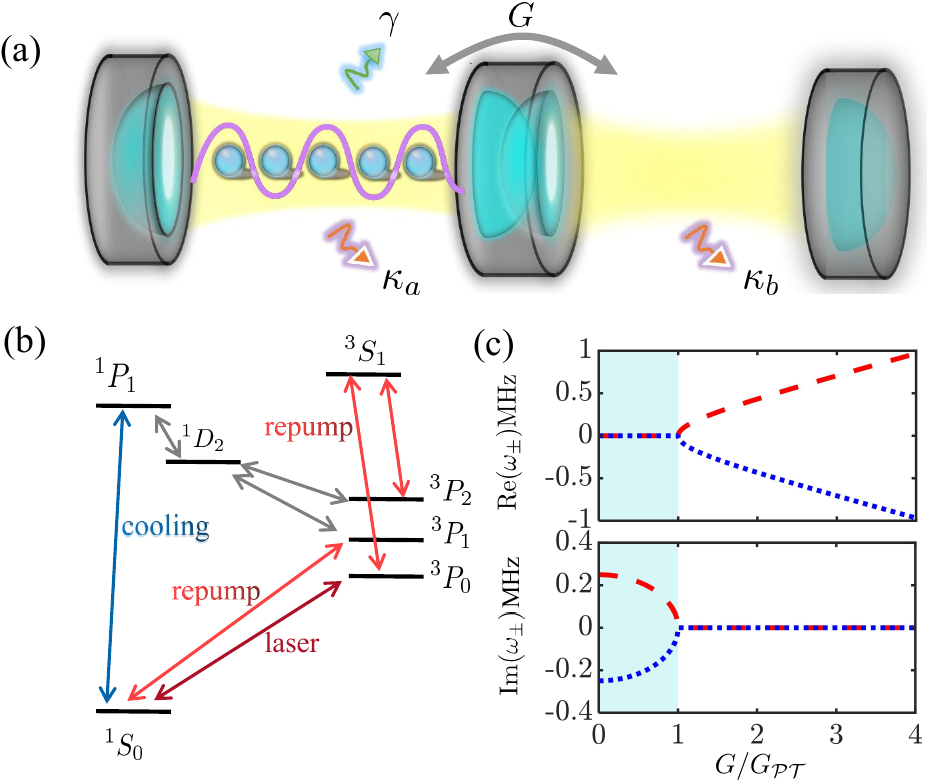}
	\caption{(a) Thousands of $^{87}{\rm Sr}$ atoms are trapped in an optical cavity coupled to an auxiliary empty cavity with coupling strength $G$. (b) The level structure of $^{87}{\rm Sr}$: the dipole-allowed $^1S_0$ to $^1P_1$ transition is used for the initial laser cooling. After excitation, not all atoms return to the ground state. A fraction of the atoms populate the $^1D_2$ state, which subsequently decays to the metastable $^3P_2$ and $^3P_1$ states. Atoms in the $^3P_1$ state can spontaneously decay to $^1S_0$ state because the spin-orbit coupling and hyperfine interactions in $^{87}{\rm Sr}$ enable weakly dipole-forbidden (doubly forbidden) transitions, such as $^1S_0$\,-\,$^3P_1$($^1S_0$\,-\,$^3P_0$). Atoms in $^3P_2$ state, which are unable to decay to $^1S_0$ state due to selection rules, are repumped to the $^3S_1$ state, where atoms undergo spontaneous emission with fixed branching ratios to the $^3P_2$, $^3P_1$ and $^3P_0$ states. An additional repumping laser is employed to transfer atoms from $^3P_0$ back to $^3S_1$ due to the relatively long lifetime of the $^3P_0$ state. This scheme continuously maintains the population of atoms in the $^3P_1$ state, from which atoms transition back to the ground state, thereby establishing a closed cycle that enhances cooling efficiency. The dipole-forbidden $^1S_0$ to $^3P_1$ transition is employed for final cooling into the optical lattice. Superradiant lasing is observed on the dipole-forbidden $^1S_0$ to $^3P_0$ transition at 698 $\mathrm{nm}$. (c) The real and imaginary parts of eigenvalues of the tunneling-coupled subsystem versus $G$. The shaded areas and other areas represent the  $\mathcal{PT}$-symmetry-breaking phase ($\mathcal{PT}$BP) and $\mathcal{PT}$-symmetry phase ($\mathcal{PT}$SP), respectively. System parameters are $\Delta_a=\Delta_b=0$, $\kappa_a/2\pi=160\,{\rm kHz}$, and $\kappa_b/2\pi=1\,{\rm kHz}$.\label{fig1}}
\end{figure}

\emph{System.}---We consider a cavity quantum electrodynamics (QED) system consisting of $N$ identical two-level atoms with transition frequency $\omega_{\sigma}$ confined in an optical cavity by an optical lattice. The optical cavity is coupled to an auxiliary empty cavity via optical tunneling with coupling strength $G$ [see Fig.\,\ref{fig1}(a)].  The system Hamiltonian reads 
\begin{eqnarray}
	H&=&\hbar\omega _{a}a^{\dag }a+\hbar\omega _{b}b^{\dag }b+\hbar G(a^{\dag}b+ab^{\dag})\nonumber\\&&+\hbar\omega_{\sigma}\sum_{i=1}^N\sigma_{i}^{+}\sigma_{i}^{-}+\hbar g\sum_{i=1}^N(a^{\dag}\sigma_i^{-}+a\sigma_{i}^{+}),\label{eq1}
\end{eqnarray}
where $a$ ($a^{\dagger}$) and $b$ ($b^{\dagger}$) are the annihilation (creation) operators of two cavity modes with frequencies $\omega_a$ and $\omega_b$, respectively. Moreover $\sigma^{-}_{i}\equiv|g\rangle\langle e|=(\sigma_{i}^{+})^\dag$ is the lowering operator for the $i$th atom, with $|e\rangle$ and $|g\rangle$ representing the two clock levels $^3P_{0}$ and $^{1}S_{0}$ in $^{87}$Sr\,\cite{Matthew A. Norcia}, as shown in Fig.\,\ref{fig1}(b), and $g$ denotes the coupling strength between individual atoms and the optical cavity mode. Based on the recent experiments regarding the normal superradiant lasers\,\cite{ M. A. Norcia, J. G. Bohnet, Matthew A. Norcia, T. Laske, S. Dubey, J. R. K. Cline} and EP\,\cite{M. Fan, Z. Li}, our model is realistic enough for experimental implementation\,\cite{supple}.
\par
To investigate the effect of the EP on superradiant lasing, we first focus on the tunneling-coupled subsystem consisting of two cavities. Taking the dissipations into consideration and applying a rotating frame with frequency  $\omega_{\sigma}$, the tunneling-coupled system can be described by an effective non-Hermitian Hamiltonian $H_{ab}^{\rm eff}/{\hbar}=(\Delta_a-i\kappa_{a}/2)a^{\dagger}a+(\Delta_b-i\kappa_{b}/2)b^{\dagger}b+G(a^{\dagger}b+ab^{\dagger})$, where $\Delta_{a(b)}=\omega_{a(b)}-\omega_{\sigma}$ denotes the detuning of cavity mode $a$ $(b)$ and atomic transition frequency. Here $\kappa_a$ and $\kappa_b$ are the decay rates of the cavity mode $a$ and  $b$, respectively. By defining $\chi=(\kappa_a+\kappa_b)/4$ and applying a gauge transformation $(a,b)^T=e^{-\chi t}(A,B)^T$\,\cite{S. K. Ozdemir}, the effective Hamiltonian in the bases of $(A,B)^T$  becomes 
\begin{gather}\label{eq3}
	\!\!\!\!H_{ab}^{\rm eff}/{\hbar}\!=\!\begin{pmatrix}\Delta_{a}+i(\kappa_{b}-\kappa_{a})/{4} & G \\ G & \Delta_{b}-i(\kappa_{b}-\kappa_{a})/{4}  \end{pmatrix}.
\end{gather}
This Hamiltonian possesses $\mathcal{PT}$ symmetry and can undergo a $\mathcal{PT}$ symmetry phase transition as $G$ crosses the EP $G_{\mathcal{PT}}$. For example,  as shown in Fig.\,\ref{fig1}(c), this occurs when $G_{\mathcal{PT}}=(\kappa_{a}-\kappa_{b})/4$ for $\Delta_{a}=\Delta_{b}=0$, at which point the eigenvalues and corresponding eigenstates coalesce\,\cite{supple}. Note that, this is a clear two-cavity-effect.
\par
\emph{EP-controlled atomic coherence.}---We take the steady-state emission of the total system into account by employing the Lindblad master equation
\begin{eqnarray}\label{eq31}
	\frac{\partial}{\partial t}\rho&=&-\frac{i}{\hbar}[H,\rho]+\kappa_{a}\mathcal{L}[a]\rho+\kappa_{b}\mathcal{L}[b]\rho
	+\gamma\sum_{i=1}^{N}\mathcal{L}[\sigma_{i}^{-}]\rho	
	\nonumber\\&&+\eta\sum_{i=1}^{N}\mathcal{L}[\sigma_{i}^{+}]\rho	
	+\gamma_{\phi}\sum_{i=1}^{N}\mathcal{L}[\sigma_{i}^{+} \sigma_{i}^{-}]\rho,
\end{eqnarray}
where $\mathcal{L}[\mathcal{O}]\rho=\mathcal{O}\rho \mathcal{O}^{\dag}-\{\mathcal{O}^{\dag}\mathcal{O},\rho\}/2$. Here $\gamma$, $\eta$, and $\gamma_{\phi}$ represent the spontaneous emission, incoherent repumping, and pure dephasing rates of individual atoms, respectively. Under the condition $\kappa_b\gg\gamma$, the empty cavity mode can be adiabatically eliminated, yielding $b=-2iGa/\kappa_b$. Consequently, the dissipative terms for the two cavity modes in the master equation simplify to $\kappa_a\mathcal{L}[a]\rho+\kappa_b\mathcal{L}[b]\rho=\kappa_{\rm eff}\mathcal{L}[a]\rho$, where the effective dissipation $\kappa_{\rm eff}=\kappa_a+4G^2/\kappa_b$. Based on this, we derive the cooperativity parameter $C=4g^2\kappa_b/[(\kappa_a\kappa_b+4G^2)\gamma]$, which quantifies the relative strength of the atom–cavity coupling compared with dissipative processes, including free-space atomic decay and cavity-field loss. The corresponding Purcell enhanced atomic emission rate is $\Gamma_{c}=C\gamma$, arising from the modified electromagnetic local density of states provided by the resonant cavity. 
\begin{figure}
	\centering
	\includegraphics[width=8.7cm]{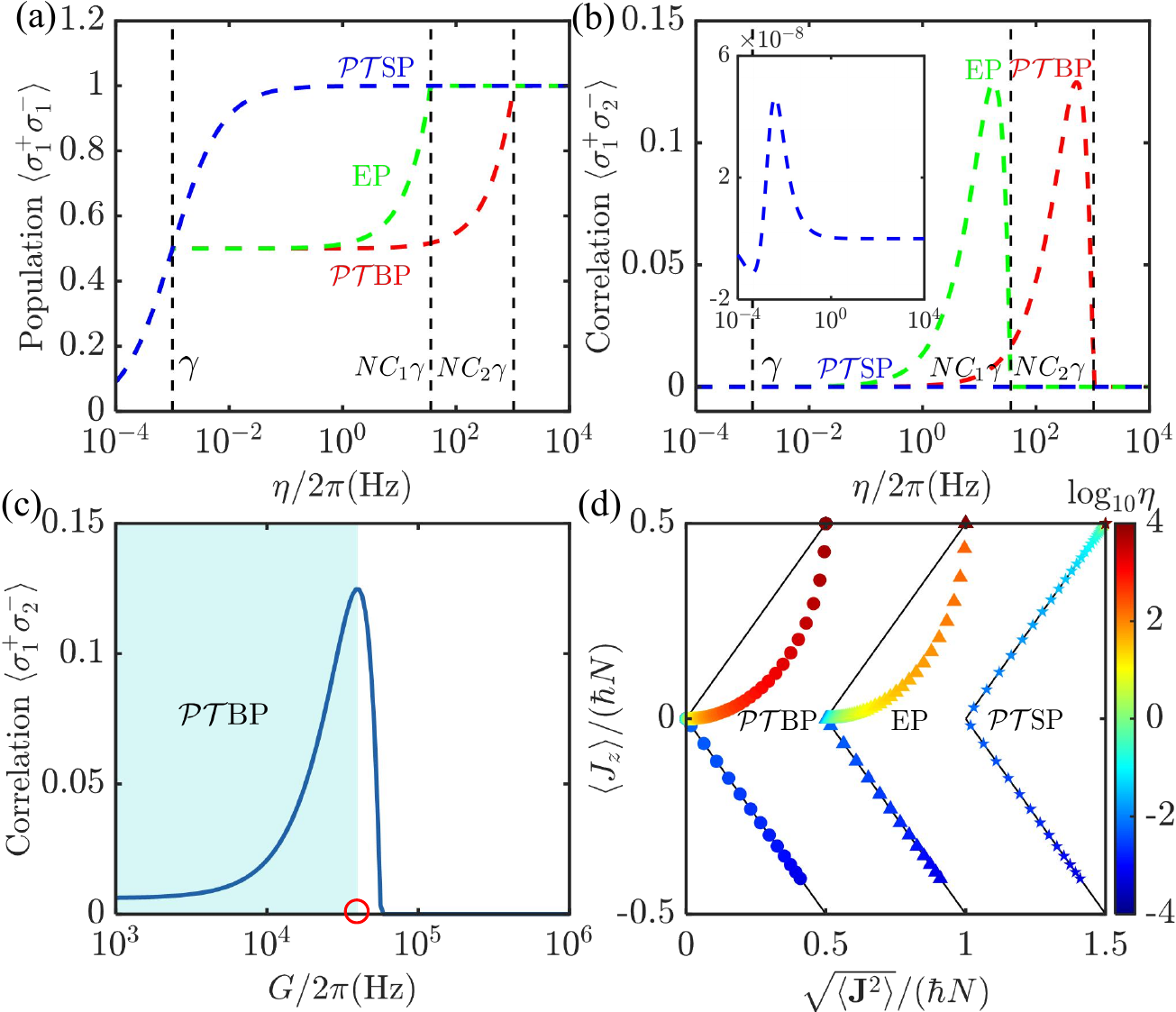}
	\caption{(a) Steady-state excited-state population of an individual atom $\langle \sigma_{1}^{+}\sigma_{1}^{-}\rangle$ and (b) Atom-atom correlation $\langle \sigma_{1}^{+} \sigma_{2}^{-}\rangle$ versus $\eta$ in the $\mathcal{PT}$BP ($G/2\pi=3.975$\,kHz, $C_2/2\pi=16.6$\,mHz), EP ($G/2\pi=39.75$\,kHz, $C_1/2\pi=0.571$\,mHz),  and (inset) $\mathcal{PT}$SP ({$G/2\pi=3975$\,kHz}). (c) Atom-atom correlation $\langle \sigma_{1}^{+} \sigma_{2}^{-}\rangle$ versus $G$ when $\eta/2\pi=18$\,Hz. The shaded region corresponds to the $\mathcal{PT}$BP. (d) Steady-state Dicke state population  with  varying $\eta$ in the $\mathcal{PT}$BP, EP, and $\mathcal{PT}$SP (left to right). The results are shifted for clarity and the thin black lines mark Dicke ladder boundaries. System parameters are $N=10^{7}$, $\Delta_a=\Delta_b=0$, $g/2\pi=2.41\,{\rm Hz}$, $\kappa_a/2\pi=160\,{\rm kHz}$, $\kappa_b/2\pi=1\,{\rm kHz}$, $\gamma/2\pi=1\,{\rm mHz}$, and $\gamma_{\phi}/2\pi=1\,{\rm mHz}$.}\label{fig2}
\end{figure}
\par
For a system with thousands of atoms, obtaining exact analytical results for dynamics is challenging. By using second-order mean-field theory\,\cite{D. Meiser} (equivalent to the cluster expansion method\,\cite{H. A. M. Leymann}) and accounting for symmetry in the expectation values related to particle exchange, we derive a closed set of equations of motion from the master equation to describe the system dynamics\,\cite{supple}. In the steady-state limit, we obtain an equation
\begin{eqnarray}\label{eq4}
	[NC\gamma\!+\!\Gamma\!-\!\frac{2NC\gamma\eta}{\eta+\gamma}\!+\!\frac{2N^{2}C^{2}\gamma^{2}}{\eta+\gamma}\langle\sigma_{1}^{+}\sigma_{2}^{-}\rangle]\langle\sigma_{1}^{+}\sigma_{2}^{-}\rangle\!=0,
\end{eqnarray}
where $\Gamma=\eta+\gamma+\gamma_{\phi}$ is the total atomic dipole relaxation rate. Here $\langle\sigma_{1}^{+}\sigma_{2}^{-}\rangle$  characterizes atom–atom correlations and serves as a key indicator of collective radiative behavior, distinguishing collective radiance from independent spontaneous emission\,\cite{A. Shankar}. In the limit $\Gamma/(NC\gamma)\rightarrow0$, this reduces to $[1-2\eta/(\eta+\gamma)+2NC\gamma/(\eta+\gamma)\langle\sigma_{1}^{+}\sigma_{2}^{-}\rangle]\langle\sigma_{1}^{+}\sigma_{2}^{-}\rangle=0$. 
At the laser threshold where the gain surpasses losses $\eta=\gamma$,  atom-atom correlations change sign,  marking the onset of collective behavior. As the pump rate increases further, the correlations shift sign again, indicating a return to the non-collective radiance regime. The upper laser threshold, or maximum pump rate in the lasing regime, is reduced to
\begin{eqnarray}
	\eta_{\rm max}=NC\gamma=\frac{4Ng^2\kappa_b}{4G^2+\kappa_a\kappa_b}.
\end{eqnarray}
This indicates that the $\mathcal{PT}$-symmetric phase transition of subsystem, decided by the system parameters $G$, $\kappa_a$, and $\kappa_b$, can affect the upper threshold of superradiant lasing regime.
\par
Futherly, Figs.\,\ref{fig2}(a) and \ref{fig2}(b) show the steady-state excited-state population $\langle \sigma_{1}^{+}\sigma_{1}^{-}\rangle$ and atom-atom correlations $\langle \sigma_{1}^{+} \sigma_{2}^{-}\rangle$ in different phase regions and the EP. The atomic population initially increases at the same rate in all three cases when $\eta<\gamma$. However, as the system reaches the threshold $\eta=\gamma$,  the three curves diverge. Beyond this threshold, the atomic population  in the $\mathcal{PT}$SP rises to a  maximum, while in the EP and $\mathcal{PT}$BP cases, it initially stabilizes on a plateau before eventually peaking at the upper threshold $\eta= \eta_{\rm max}$ [see Fig.\,\ref{fig2}(a)].  The atom-atom correlation shown in Fig.\,\ref{fig2}(b) transitions from negative, indicating suppressed collective radiance, to positive, signifying enhanced collective radiance, as $\eta$ increases past $\eta = \gamma$. This originally comes from the indirect atom-atom interactions induced by cavity. The correlation then peaks and sharply drops to zero at the upper threshold $\eta = \eta_{\rm max}$, reflecting a transition to non-collective radiance. The comparison of the main and inset parts in Fig.\,\ref{fig2}(b) demonstrates that both the EP and $\mathcal{PT}$BP phases significantly enhance atom-atom correlations in the superradiance regime. Finally, Fig.\,\ref{fig2}(c) illustrates the impact of the system parameter $G$ on the atom-atom correlation, and this enhancement of  atomic correlation at EP also reflects a significant cavity effect.
\begin{figure}
	\centering
	\includegraphics[width=8.8cm]{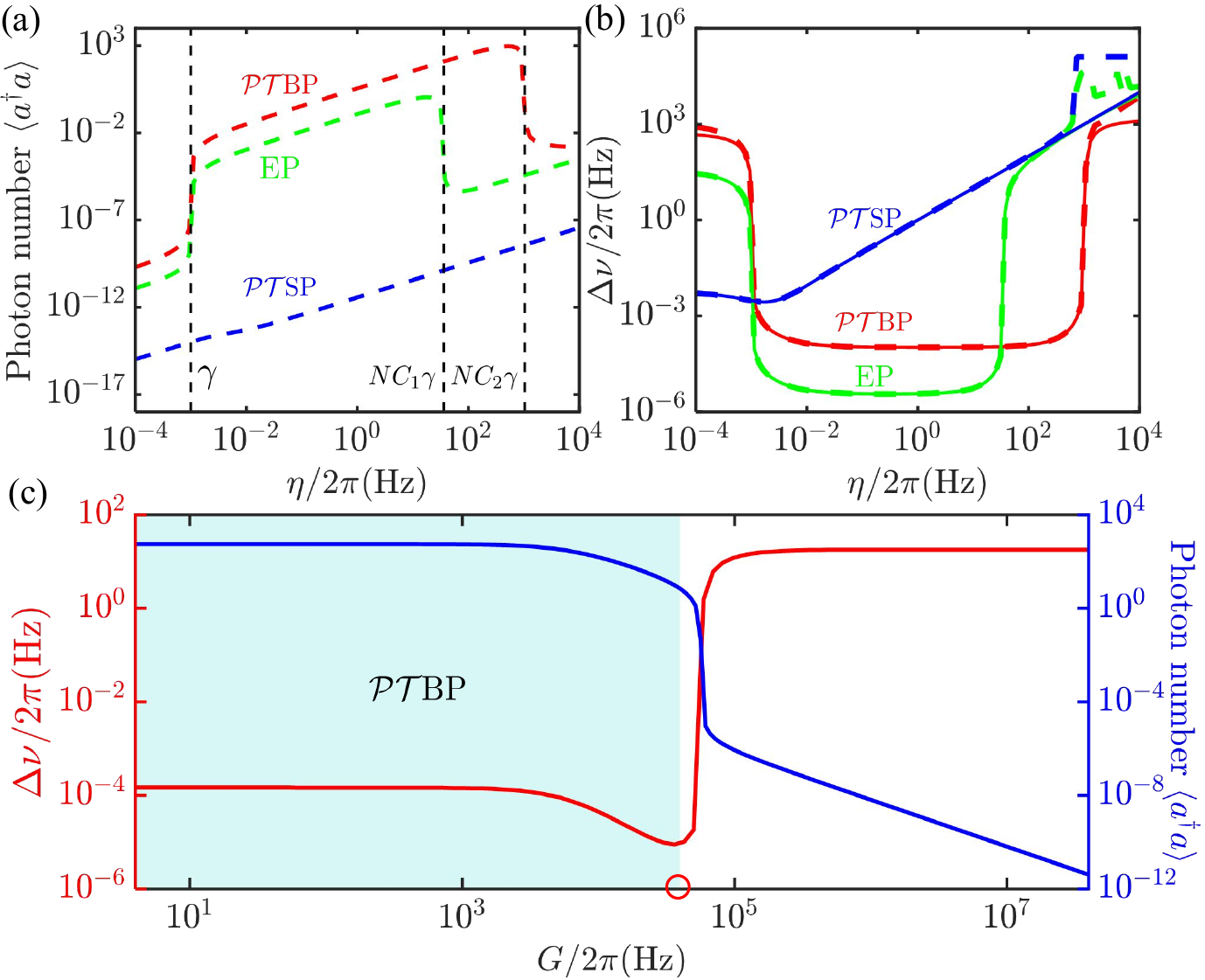}
	\caption{(a) Steady-state intracavity photon number $\langle a^{\dagger}a\rangle$ and (b) emission linewidth $\Delta\nu$ versus $\eta$ in the $\mathcal{PT}$BP, EP,  and $\mathcal{PT}$SP.  The solid lines in (b) correspond to the analytical approximate solutions from Eq.\,(\ref{eq5}). (c)  $\langle a^{\dagger}a\rangle$ and  $\Delta\nu$ versus $G$ for $\eta/2\pi=18$\,Hz. The shaded region corresponds to the $\mathcal{PT}$BP. Other system parameters are the same as in Fig.\,\ref{fig2}. }\label{fig4}
\end{figure}
\par
Figure\,\ref{fig2}(d) shows the steady-state Dicke state $|J,M\rangle$\,\cite{R. H. Dicke} population in different phase regions and the EP. Here $|J,M\rangle$ is the joint eigenstate of the operators $\bold{J}^2$ and ${J}_z$, where ${\bold{J}}^2/\hbar^2|J,M\rangle=J(J+1)|J,M\rangle$ and $J_z/\hbar|J,M\rangle=M|J,M\rangle$ with $J=0,1,2,\dots,N/2$ (for an even number $N$) and $M=-J,-J+1,\dots,J-1,J$. We have defined the operators $J_j/\hbar=\sum_{i=1}^N \sigma_i^j/2\,(j=x,y,z)$, where $\sigma_i^j$ are the Pauli operators for the $i\mathrm{th}$ atom. Using the second-order mean-field approach, we calculate the Dicke state  population, with $\langle J_z \rangle/\hbar =N(\langle\sigma_{1}^{+}\sigma_{1}^{-}\rangle-1/2)$ and $\sqrt{\langle \bold{J}^2\rangle}/\hbar =\sqrt{\frac{3N}{4}\!+\!N(N\!-\!1)(\langle\sigma_{1}^{+}\sigma_{2}^{-}\rangle\!+\!\langle\sigma_{1}^{+}\sigma_{1}^{-} \sigma_{2}^{+}\sigma_{2}^{-}\rangle\!-\!\langle\sigma_{1}^{+}\sigma_{1}^{-}\rangle\!+\!\frac{1}{4})}$. We observe that,  within the given parameter range, the system is initially at the subradiance regime in all three cases. In this scenario, $\eta$ is small and the system operates in the low-excitation regime, where the average number of excited atoms $M+N/2$ is much smaller than the total atom number $N$\,\cite{N. Lambert}. In the closed system, the atomic ensemble can be regarded as one bright mode and $N-1$ dark modes, with only the bright mode coupling to the cavity mode\,\cite{T. E. Li,N. Moiseyev, L. P. Lindoy}. In contrast, in the open system considered here, dephasing induces coupling between the bright and dark modes, resulting in an indirect interaction between the dark modes and cavity mode. As $\eta$ increases, the system transitions from a subradiance state to superradiance state with low symmetry and then ascends the Dicke ladder towards higher-symmetry superradiance states, ultimately reaching the fully excited states\,\cite{N. Shammah}. In this scenario, the two-level atoms can no longer be approximated as bosons, which prevents the atomic ensemble being described in terms of two uncoupled bright and dark modes\,\cite{supple}. The atomic ensemble collectively couples to the cavity mode. Notably, there is a sharp transition that the population doesn't ascend along the Dicke state boundary as $\eta$ increases in the EP and $\mathcal{PT}$BP cases, which is absent in the $\mathcal{PT}$SP case. This behavior aligns with the results in Fig.\,\ref{fig2}(a) and \ref{fig2}(b), indicating that  superradiance mainly comes from atomic coherence. As the system approaches the fully excited state, the  atom-atom correlation $\langle\sigma_{1}^{+}\sigma_{2}^{-}\rangle\rightarrow0$ [see Fig.\,\ref{fig2}(b)].
\par
\emph{EP-induced ultranarrow linewidth of superradiant lasing.}---In the cavity QED system, the excited atomic ensemble releases energy by emitting photons into the cavity,  allowing us to observe superradiant lasing effects by monitoring the intracavity photon number. Similar to conventional cavity QED systems \cite{K. Debnath}, here the proposed system with a $\mathcal{PT}$-symmetric subsystem also undergoes  radiance transitions from subradiance to superradiance and finally to non-collective radiance as the pump rate $\eta$ increases. However, in our proposal, the intracavity photon number $\langle a^{\dagger}a\rangle$ exhibits sharp increases and decreases  at the lasing threshold points $\eta=\gamma, \eta_{\rm max}$ in the EP and $\mathcal{PT}$BP, [see Fig.\,\ref{fig4}(a)].  The value of  $\langle a^{\dagger}a\rangle$ can reach  the order of $10^3$ in the $\mathcal{PT}$BP and around $10$ in EP. This notable enhancement in the superradiant lasing regime is attributed to the substantial boost in the atom-atom correlations within the EP and $\mathcal{PT}$BP [see Fig.\,\ref{fig2}(b)]. 
\par
\begin{figure}
	\centering
	\includegraphics[width=8.0cm]{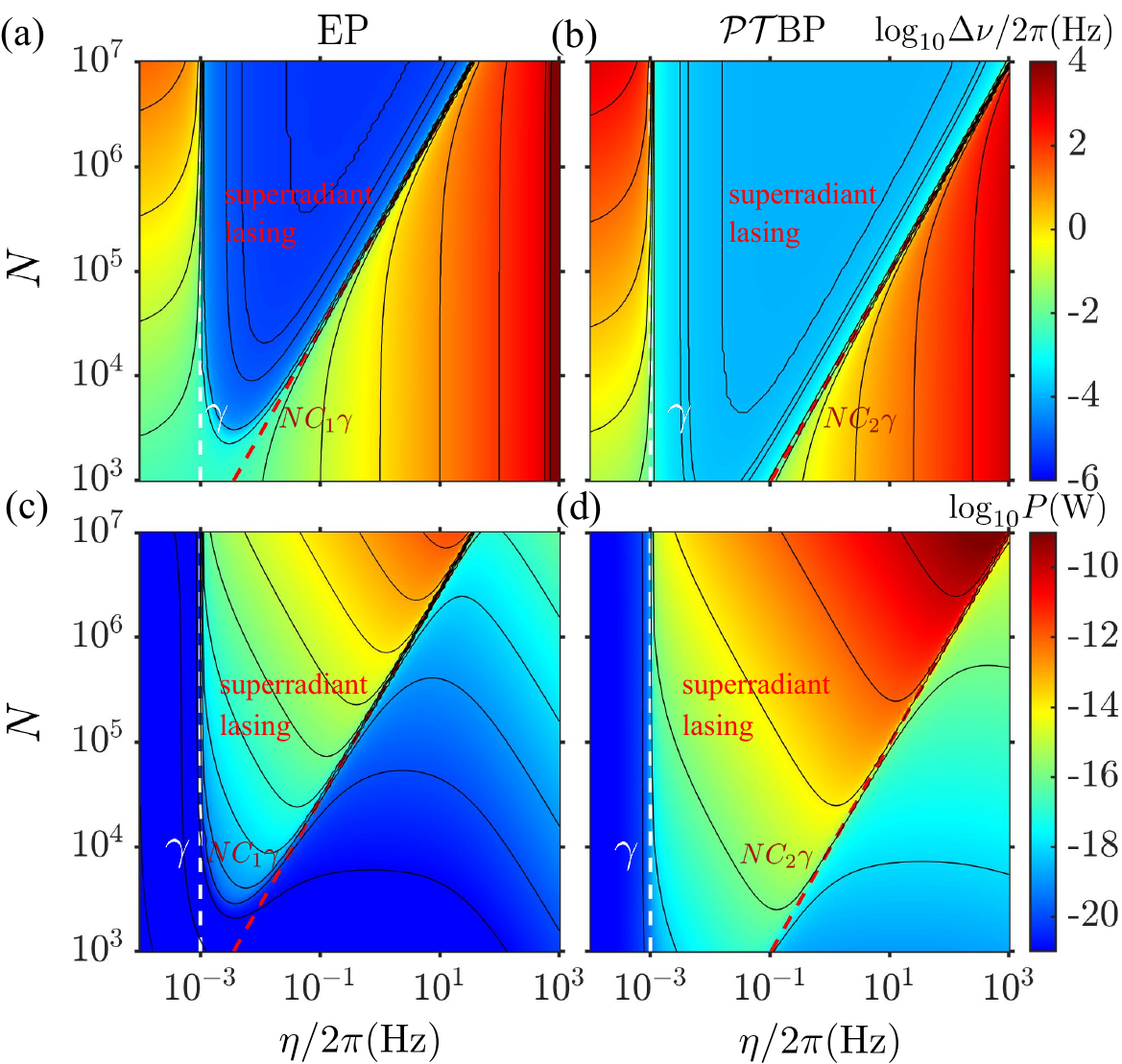}
	\caption{(a,b) Emission linewidth $\Delta\nu$ and (c,d) output power $P$ in the steady-state versus $\eta$ and $N$ for (a,c) EP and (b,d) $\mathcal{PT}$BP.  The  white and red dashed lines denote the thresholds for the superradiant lasing. Other system parameters are the same as in Fig.\,\ref{fig2}.}\label{fig5}
\end{figure}
Besides photon number,  linewidth is a crucial metric for evaluating superradiant lasing. By solving Eq.\,(\ref{eq31}), we obtain the steady-state emission spectrum of the light field. Figure \ref{fig4}(b) shows the emission linewidth $\Delta\nu$ for the system in three different regions. The linewidth $\Delta\nu$ drops sharply at the lasing threshold $\eta=\gamma$ and remains ultranarrow in the superradiant lasing regime for both the $\mathcal{PT}$BP and EP, while it increases almost steadily in the $\mathcal{PT}$SP. The minimum linewidth reaches approximately $2\pi\times10^{-4}$ Hz in $\mathcal{PT}$BP and $2\pi\times10^{-6}$ Hz in EP, where atom-atom correlations peak, significantly amplifying collective effects. The linewidth  in the EP is three orders of magnitude narrower than in systems without $\mathcal{PT}$-symmetry. This ultranarrow linewidth is even smaller than the individual atomic decay rate $\gamma$.  Figure\,\ref{fig4}(c) further shows the effects of the system parameter $G$ on the linewidth, with emission linewidth  reaching a minimum in the EP while maintaining a relatively high photon number. Additionally, we calculate the emission spectrum in the EP by coupling the cavity output to a filter cavity using second-order  mean-field theory\,\cite{supple}. These results also confirm that superradiant lasing with ultranarrow linewidth is achievable in the EP. This will motivate experimentalists to pursue the development of a new type of optical clock that would be more stable than current ones\,\cite{supple}. To characterize EP-controlled superradiant lasing more intuitively, we derive an analytical expression for the emission linewidth based on the quantum regression theorem and second-order mean-field approximation\,\cite{G. Dong} 
\begin{eqnarray}
	\!\!\!\Delta \nu&\!\!=\!\!&\frac{\Gamma\!-\!2\Gamma_{c}\langle J_z \rangle/\hbar}{1\!+\!(\kappa_{a}+\kappa_{b})\Gamma/(4G^{2}+\kappa_{a}\kappa_{b})\!-\!2\langle J_z \rangle\Gamma_{c}/(\hbar\kappa_{b})} \label{eq5}. 
\end{eqnarray}
This expression depends on the system parameters $G$, $\kappa_a$, and $\kappa_b$, and then it clearly demonstrates that the $\mathcal{PT}$ symmetry property of the system will influence the emission linewidth significantly. At the EP, the linewidth can be  further simplified to  $\Delta \nu=(\Gamma-2\Gamma_c\langle J_z \rangle / \hbar)/[1+4\Gamma/(\kappa_a+\kappa_b)-2\langle J_z \rangle \Gamma_c/(\hbar\kappa_b)]$, and for a given $\eta$, it is fundamentally determined by the cavity losses $\kappa_a$, $\kappa_b$  and the atom-cavity coupling strength $g$\,\cite{supple}. Figure \ref{fig4}(b) illustrates that the analytical emission linewidths are excellent agreement with the fully numerical simulations, validating our approximation approach.
\par
Conventional lasers rely on optical coherence achieved through stimulated emission, with a fundamental linewidth constrained by the Schawlow–Townes limit\,\cite{A. L. Schawlow}, which is inversely proportional to the photon number in the cavity. In contrast, superradiant lasing, where the atomic ensemble acts as the gain medium similar to conventional lasers, relies on atomic coherence to achieve high-intensity output. Consequently, superradiant lasing is highly dependent on the atom number $N$. Figure\,\ref{fig5} illustrates the impact of  $N$ and  $\eta$ on both emission power $P=\hbar\kappa\omega_a\langle a^{\dagger}a \rangle$ and linewidth $\Delta \nu$ in the $\mathcal{PT}$BP and EP. In the superradiant lasing regime—marked by two dashed lines—changes in $N$ significantly influence emission power but have minimal impact on the linewidth, which remains extremely narrow at approximately $2\pi\times10^{-4}$ Hz in the $\mathcal{PT}$BP and $2\pi\times10^{-6}$ Hz in the EP.  Thus, the superradiant lasing with ultranarrow linewidth and high-power level can be observed in our system. Moreover, increasing $N$ also broadens the  range of superradiant lasing, as a larger atom number requires a stronger incoherent pumping rate $\eta$ to drive the system to the fully excited state of the atomic ensemble, going to uncorrelated radiance regime. In realistic systems, the fluctuations of cavity frequency\,\cite{J. G. Bohnet} and atom number\,\cite{Matthew A. Norcia,T. Laske, S. Dubey} can both affect the resulting linewidth\,\cite{supple}. To achieve an ultranarrow linewidth, these fluctuations must remain sufficiently low to decrease the broadening effects.

\par
\emph{Conclusions}.---We have investigated superradiant lasing in a cavity QED system featuring a controllable $\mathcal{PT}$-symmetry phase transition. Our findings demonstrate that the $\mathcal{PT}$-symmetry phase transition has a significant effect on superradiant lasing, enabling ultranarrow linewidth lasing in the EP. The linewidth of superradiant lasing  in EP is three orders of magnitude narrower than in systems without $\mathcal{PT}$-symmetry, which could enhance  stability and precision in applications like atomic clocks. This work opens up new directions at the intersection of the study of symmetry and the field of quantum sensing with ultra-narrow linewidth, and can be extended to multimode non-Hermitian systems to explore the effects of higher-order EPs\,\cite{H. Hodae, S. Wang} on superradiant lasing.
\par

{\it Acknowledgments}—We sincerely thank Zhi-Guang Lu and Guoqing Tian for their valuable suggestions. This work is supported by the Quantum Science and Technology-National Science and Technology Major Project (Grant No.\,2024ZD0301000), the National Science Fund for Distinguished Young Scholars of China (Grant No.\,12425502), the National Natural Science Foundation of China (Grant No.\,12547108, No.\,12574397), the Sichuan Science and Technology Program (Grant No.\,2025ZNSFSC0057), the National Key Research and Development Program of China (Grant No.\,2021YFA1400700), and the Fundamental Research Funds for the Central Universities (Grant No.\,2024BRA001). F.\,N. is supported in part by the Japan Science and Technology Agency (JST) [via the CREST Quantum Frontiers program Grant No.\,JPMJCR24I2, the Quantum Leap Flagship Program (Q-LEAP), and the Moonshot R\&D Grant No.\,JPMJMS2061], and the Office of Naval Research (ONR) Global (via Grant No.\,N62909-23-1-2074). The computation is completed in the HPC Platform of Huazhong University of Science and Technology.

{\it Data availability}—The data that support the findings of this Letter are not publicly available. The data are available from the authors upon reasonable request.

\onecolumngrid
\clearpage
\setcounter{equation}{0}
\setcounter{figure}{0}
\setcounter{table}{0}
\setcounter{page}{7}
\setcounter{section}{0}
\makeatletter
\renewcommand{\theequation}{S\arabic{equation}}
\renewcommand{\thefigure}{S\arabic{figure}}
\renewcommand{\bibnumfmt}[1]{[S#1]}
\renewcommand{\citenumfont}[1]{S#1}
\setcounter{secnumdepth}{7} \makeatletter

\begin{center}
        \textbf{Supplemental Material for ``Exceptional Point Superradiant Lasing With Ultranarrow Linewidth"}
\end{center}

\begin{center}
	Min Du$^{1}$, Qian Bin$^{2,*}$, Qing-Yang Qiu, Franco Nori$^{3,4,5}$, Xin-You L\"{u}$^{1,\dagger}$
\end{center}
\begin{center}
	\begin{minipage}[]{16cm}
		\small{\it
			\centering $^{1}$ School of Physics and Institute for Quantum Science and Engineering, Huazhong University of Science and Technology and Wuhan Institute of Quantum Technology, Wuhan 430074, China\\
			$^{2}$ College of Physics, Sichuan University, Chengdu 610065, China\\
			$^{3}$ Theoretical Quantum Physics Laboratory, Cluster for Pioneering Research, RIKEN, Wako-shi, Saitama 351-0198, Japan\\
			$^{4}$ Quantum Information Physics Theory Research Team, Quantum Computing Center, RIKEN, Wakoshi, Saitama, 351-0198, Japan\\
			$^{5}$ Physics Department, The University of Michigan, Ann Arbor, Michigan 48109-1040, USA}
	\end{minipage}
\end{center}
	
\title{Supplemental Material for ``Nonreciprocal Bundle Emissions of Quantum Entangled Pairs"}
\date{\today}

\title{Supplemental Material for ``Nonreciprocal Bundle Emissions of Quantum Entangled Pairs"}
\author{Qian Bin}
\affiliation{School of Physics and Institute for Quantum Science and Engineering, Huazhong University of Science and Technology,
and Wuhan Institute of Quantum Technology, Wuhan 430074, China}

\author{Hui Jing}
\affiliation{Key Laboratory of Low-Dimensional Quantum Structures and Quantum Control of Ministry of Education, Department of Physics and Synergetic Innovation Center for Quantum Effects and Applications, Hunan Normal University, Changsha 410081, China}

\author{Ying Wu}
\affiliation{School of Physics and Institute for Quantum Science and Engineering, Huazhong University of Science and Technology,
and Wuhan Institute of Quantum Technology, Wuhan 430074, China}

\author{Franco Nori}
\affiliation{Theoretical Quantum Physics Laboratory, Cluster for Pioneering Research, RIKEN, Wakoshi, Saitama 351-0198, Japan}
\affiliation{Center for Quantum Computing, RIKEN, Wakoshi, Saitama 351-0198, Japan}
\affiliation{Physics Department, The University of Michigan, Ann Arbor, Michigan 48109-1040, USA}

\author{Xin-You L\"{u}}
\email{xinyoulu@hust.edu.cn}
\affiliation{School of Physics and Institute for Quantum Science and Engineering, Huazhong University of Science and Technology,
and Wuhan Institute of Quantum Technology, Wuhan 430074, China}

\date{\today}
\maketitle

This Supplemental Material contains five parts: I. Derivation of eigenvectors and eigenvalues. II. Second-order mean-field equations and derivation of analytical expressions in the steady-state limit.
III. Dicke-state population and atom-atom correlation in the low- and high-excitation regimes. IV. Approximate analytical expression of spectrum linewidth obtained by the quantum regression theorem. V. Calculation of spectrum with a filter cavity.  VI. Frequency pulling by cavity detuning and atom number fluctuations by technical noises. VII. Experimental feasibility and stability analysis of different optical atomic clocks.

\section{Derivation of eigenvectors and eigenvalues}\label{I}
From the main text, we derive that the effective non-Hermitian Hamiltonian describing the dissipative, tunneling-coupled subsystem after a gauge transformation is given by
\begin{equation}
	H_{\rm eff}^{'}=
	{\hbar}\left( \begin{array}{ccc}
		\Delta_{a}+i(\kappa_{b}-\kappa_{a})/4& G\\
		G & \Delta_{b}-i(\kappa_{b}-\kappa_{a})/4
	\end{array} 
	\right ),
\end{equation}
where the eigenvalues are ${\hbar}\omega_{\pm}=\pm{\hbar}\sqrt{G^2-(\kappa_{a}-\kappa_{b})^2/16}$ under the condition of assuming resonance between the cavity mode and atom frequencies, i.e., $\Delta_a=\Delta_b=0$. 

In the case $G>(\kappa_{a}-\kappa_{b})/4$, the subsystem resides in the exact $\mathcal{PT}$ symmetry phase,  and the eigenvalues simplify to $\hbar\omega_{\pm}=\pm \hbar G \cos\phi$, where $\phi=\arcsin[(\kappa_{a}-\kappa_{b})/4G]$. The eigenvector for ${\hbar}\omega_{+}$ is solved by 
\begin{equation}
	(H_{\rm eff}^{'}-\omega_{+}\mathbb{I})/{\hbar}\to \begin{pmatrix} -i \sin\phi- \cos\phi & 1 \\ 1 & i\sin\phi-\cos\phi	\end{pmatrix} \to \begin{pmatrix} -e^{i\phi} & 1 \\ 1 & -e^{-i\phi}\end{pmatrix}\to \begin{pmatrix} -e^{i\phi} & 1 \\ 0 & 0\end{pmatrix}, 
\end{equation}
yielding $|\omega_{+}\rangle=\frac{1}{{\sqrt{2}}}(1,e^{i\phi})^{T}$, where $\mathbb{I}$ is an identity matrix. Similarly, the eigenvector for ${\hbar} \omega_{-}$ is obtained as  
\begin{equation}
	(H_{\rm eff}^{'}-\omega_{-}\mathbb{I})/{\hbar}\to \begin{pmatrix} -i \sin\phi+ \cos\phi & 1 \\ 1 & i\sin\phi+\cos\phi	\end{pmatrix} \to \begin{pmatrix} e^{-i\phi} & 1 \\ 1 & e^{i\phi}\end{pmatrix}\to \begin{pmatrix} e^{-i\phi} & 1 \\ 0 & 0\end{pmatrix}, 
\end{equation}
giving $|\omega_-\rangle=\frac{1}{{\sqrt{2}}}(1,-e^{-i\phi})^{T}$. The inner product $\langle\omega_{+}|\omega_{-}\rangle=0$ confirms the orthogonality of the eigenvectors $\{|\omega_{+}\rangle,|\omega_{-}\rangle\}$ in the $\mathcal{PT}$ symmetric phase.

In the case $G<(\kappa_{a}-\kappa_{b})/4$, the system is in the exact $\mathcal{PT}$ symmetry-breaking  phase, and the eigenvalues simplify to ${\hbar}\omega_{\pm}=\pm i {\hbar}G\sinh\phi$, where $\phi={\rm arccosh}[(\kappa_{a}-\kappa_{b})/4G]$. The eigenvector for ${\hbar}\omega_{+}$ is solved by 
\begin{equation}
	(H_{\rm eff}^{'}-\omega_{+}\mathbb{I})/{\hbar}\to \begin{pmatrix} -i \cosh\phi-i\sinh \phi & 1 \\ 1 & i\cosh\phi-i \sinh\phi	\end{pmatrix} \to \begin{pmatrix} -ie^{\phi} & 1 \\ 1 & ie^{-\phi}\end{pmatrix}\to \begin{pmatrix} -ie^{\phi} & 1 \\ 0 & 0\end{pmatrix},
\end{equation}
yielding $|\omega_{+}\rangle=\frac{1}{{\sqrt{2}}}(1,ie^{\phi})^{T}$. Similarly, the eigenvector for ${\hbar}\omega_{-}$ is solved by 
\begin{equation}
	(H_{\rm eff}^{'}-\omega_{-}\mathbb{I})/{\hbar}\to \begin{pmatrix} -i \cosh\phi+ i \sinh\phi & 1 \\ 1 & i\cosh\phi+i \sinh\phi	\end{pmatrix} \to \begin{pmatrix} -i e^{-\phi} & 1 \\ 1 & i e^{\phi}\end{pmatrix}\to \begin{pmatrix} -ie^{-\phi} & 1 \\ 0 & 0\end{pmatrix},
\end{equation}
giving $|\omega_-\rangle=\frac{1}{{\sqrt{2}}}(1,ie^{-\phi})^{T}$. The inner product $\langle\omega_{+}|\omega_{-}\rangle=0$ confirms that  $\{|\omega_{+}\rangle,|\omega_{-}\rangle\}$ are also orthogonal in the $\mathcal{PT}$ broken phase.

At the exceptional point (EP), i.e.,  $G=(\kappa_{a}-\kappa_{b})/4$, the eigenvalues coalesce to ${\hbar}\omega_{\pm}=0$,  and the corresponding eigenvectors converge to $\frac{1}{{\sqrt{2}}}(1,i)^{T}$, signifying the simultaneous coalescence of eigenvalues and eigenvectors at EP.

\section{Second-order mean-field equations and derivation of analytical expressions in the steady-state limit} \label{II}

In this section, we show how to solve the master equation using second-order mean-field theory\,\cite{SD. Meiser},  equivalent to the cluster expansion approach\,\cite{SH. A. M. Leymann}, and derive analytical expressions in the steady-state limit\,\cite{SG. Dong}. The method captures quantum correlations without relying on the assumption of spontaneous symmetry breaking required in first-order mean-field theory, which is essential for describing steady-state superradiant lasing. 

Using the master equation presented in the main text, we can derive the expectation values $\langle o \rangle=\mathrm{tr}\{o\rho\}$ of any observable $o$ of interest.  To truncate the hierarchy of expectation value involving high-order operators, we apply a third-order cumulant expansion to approximate the expectation values of three operators products. For example, we have $\langle\sigma_{1}^{+}\sigma_{1}^{-}a^{\dag}a\rangle=\langle\sigma_{1}^{+}\sigma_{1}^{-}\rangle \langle a^{\dag}a\rangle+\langle a^{\dag}\rangle \langle \sigma_{1}^{+}\sigma_{1}^{-}a\rangle+\langle a \rangle \langle \sigma_{1}^{+}\sigma_{1}^{-}a^{\dag}\rangle-2\langle a \rangle \langle \sigma_{1}^{+}\sigma_{1}^{-}\rangle \langle a^{\dag}\rangle$.  Since no coherent laser field is injected and the laser field lacks initial optical coherence, the field amplitude  $\langle a \rangle$ and polarization $\langle \sigma_1^{-} \rangle$ are always zero, yielding $\langle\sigma_{1}^{+}\sigma_{1}^{-}a^{\dag}a\rangle\approx\langle\sigma_{1}^{+}\sigma_{1}^{-}\rangle \langle a^{\dag}a\rangle$.                                                                                                            This approach preserves high-order correlations and has been applied to study steady-state superradiance\,\cite{SM. J. Holland}.

From the master equation in the main text, we derive the equations for the mean intracavity photon number 
\begin{eqnarray}
	\frac{d}{dt}\langle a^{\dagger}a\rangle&=&2G \mathrm{Im}\langle a^{\dagger}b\rangle+2gN\mathrm{Im}\langle a^{\dagger}\sigma_{1}^{-}\rangle-\kappa_{a}\langle a^{\dagger}a\rangle,\\
	\frac{d}{dt}\langle b^{\dagger}b\rangle&=&-2G\mathrm{Im}\langle a^{\dagger}b\rangle-\kappa_{b}\langle b^{\dagger}b\rangle,
\end{eqnarray}
which include the correlation terms $\langle a^{\dagger}b\rangle$ and $\langle a^{\dagger}\sigma_{1}^{-}\rangle$. Assuming all atoms are identical,  the sum over spins $\Sigma_{i=1}^N$ can be replaced with $N$, we have
\begin{eqnarray}
	\frac{d}{dt}\langle a^{\dagger}b\rangle&=&-\frac{1}{2}(\kappa_{a}+\kappa_{b})\langle a^{\dagger}b\rangle+iG(\langle b^{\dagger}b\rangle-\langle a^{\dagger}a\rangle)+i(\Delta_{a}-\Delta_{b})\langle a^{\dagger}b\rangle+iNg\langle b\sigma_{1}^{+}\rangle,\label{S8}\\
	\frac{d}{dt}\langle a^{\dagger}\sigma_{1}^{-}\rangle&=&-\frac{1}{2}(\Gamma +\kappa_{a})\langle a^{\dagger}\sigma_{1}^{-}\rangle+ig\langle\sigma_{1}^{+}\sigma_{1}^{-}\rangle+iG\langle b^{\dagger}\sigma_{1}^{-}\rangle-ig\langle a^{\dagger}a\rangle\nonumber\\&&+i\Delta_{a}\langle a^{\dagger}\sigma_{1}^{-}\rangle+ig(N-1)\langle\sigma_{1}^{+}\sigma_{2}^{-}\rangle+2ig\langle\sigma_{1}^{+}\sigma_{1}^{-}\rangle\langle a^{\dagger}a\rangle,\label{S9}\\
	\frac{d}{dt}\langle b\sigma_{1}^{+}\rangle&=&-\frac{1}{2}(\Gamma+\kappa_{b})\langle b\sigma_{1}^{+}\rangle+ig\langle a^{\dagger}b\rangle-i\Delta_{b}\langle b\sigma_{1}^{+}\rangle-iG\langle a\sigma_{1}^{+}\rangle-2ig\langle\sigma_{1}^{+}\sigma_{1}^{-}\rangle\langle a^{\dagger}b\rangle,\label{S10}
\end{eqnarray}
where $\Gamma=\gamma+\eta+\gamma_{\phi}$. The factor $N-1$ results from summing over all  atom pairs. The population of excited states $\langle\sigma_{1}^{+}\sigma_{1}^{-}\rangle$ and atom-atom correlation $\langle\sigma_{1}^{+}\sigma_{2}^{-}\rangle$ can be given by
\begin{eqnarray}
	\frac{d}{dt}\langle\sigma_{1}^{+}\sigma_{1}^{-}\rangle&=&\eta-(\gamma+\eta)\langle\sigma_{1}^{+}\sigma_{1}^{-}\rangle-2g {\rm Im}\langle a^{\dagger}\sigma_{1}^{-}\rangle,\label{S11}
\end{eqnarray}
\begin{eqnarray}
	\frac{d}{dt}\langle\sigma_{1}^{+}\sigma_{2}^{-}\rangle&=&-\Gamma\langle\sigma_{1}^{+}\sigma_{2}^{-}\rangle-2g{\rm Im}\langle a^{\dagger}\sigma_{1}^{-}\rangle+4g {\rm Im}(\langle\sigma_{1}^{+}\sigma_{1}^{-}\rangle\langle a^{\dagger}\sigma_{1}^{-}\rangle\text{),}\label{S12}
\end{eqnarray}
and for the high-order atom-atom correlation $\langle\sigma_{1}^{+}\sigma_{1}^{-}\sigma_{2}^{+}\sigma_{2}^{-}\rangle$, we have
\begin{eqnarray}
	\frac{d}{dt}\langle\sigma_{1}^{+}\sigma_{1}^{-}\sigma_{2}^{+}\sigma_{2}^{-}\rangle&=&-2\gamma\langle\sigma_{1}^{+}\sigma_{1}^{-}\sigma_{2}^{+}\sigma_{2}^{-}\rangle+2\eta(\langle\sigma_{1}^{+}\sigma_{1}^{-}\rangle-\langle\sigma_{1}^{+}\sigma_{1}^{-}\sigma_{2}^{+}\sigma_{2}^{-}\rangle)-4g\langle\sigma_{1}^{+}\sigma_{1}^{-}\rangle {\rm Im}\langle a^{\dagger}\sigma_{1}^{-}\rangle.
\end{eqnarray} 
Next, we derive analytical expressions for the atom-atom correlation $\langle\sigma_{1}^{+}\sigma_{2}^{-}\rangle$ and the excited-state population $\langle\sigma_{1}^{+}\sigma_{1}^{-}\rangle$ by neglecting all non-collective terms and dissipation terms except $\kappa_a$,\,$\kappa_b$ in Eq.\,(\ref{S9}) Eq.\,(\ref{S10}) and approximating $N-1\approx N$.  In the steady-state limit, 
\begin{eqnarray}
	\langle b^{\dagger}\sigma_{1}^{-}\rangle&=&2iG\kappa_{b}^{-1}\langle a^{\dagger}\sigma_{1}^{-}\rangle,\nonumber\\
	\langle a^{\dagger}\sigma_{1}^{-} \rangle&=&2iNg\kappa_{b}(4G^{2}+\kappa_{a}\kappa_{b})^{-1}\langle\sigma_{1}^{+}\sigma_{2}^{-}\rangle,\label{S14}
\end{eqnarray} 
indicating that the atomic-field relative phase aligns with the  macroscopic atomic dipole phase. Substituting above equations Eq.\,(\ref{S14}) into Eq.\,(\ref{S11}), we obtain the excited-state population
\begin{eqnarray}
	\langle\sigma_{1}^{+}\sigma_{1}^{-}\rangle&=&\frac{\eta}{\eta+\gamma}-\frac{4Ng^{2}\kappa_{b}(4G^{2}+\kappa_{a}\kappa_{b})^{-1}}{\eta+\gamma}\langle\sigma_{1}^{+}\sigma_{2}^{-}\rangle.\label{S15}
\end{eqnarray} 
In order to derive Eq.\,(4) in the main text, we use the expression along with $\langle a^{\dagger}\sigma_{1}^{-} \rangle$ for Eq.\,(\ref{S12}), and derive the approximate equation in the steady-state limit
\begin{eqnarray}
	[NC\gamma+\Gamma-\frac{2NC\gamma\eta}{\eta+\gamma}+\frac{2N^{2}C^{2}\gamma^{2}}{\eta+\gamma}\langle\sigma_{1}^{+}\sigma_{2}^{-}\rangle]\langle\sigma_{1}^{+}\sigma_{2}^{-}\rangle=0,
\end{eqnarray}
where the cooperativity parameter $C=4g^2\kappa_b/[(\kappa_a\kappa_b+4G^2)\gamma]$.
Solving above equation gives an analytical expression for  the spin-spin correlation
\begin{eqnarray}
	\langle\sigma_{1}^{+}\sigma_{2}^{-}\rangle&=&\frac{-(NC\gamma+\Gamma)(\eta+\gamma)+2NC\gamma\eta}{2N^2C^2\gamma^2}.
\end{eqnarray}
Substituting the above equation into Eq.\,(\ref{S15}), we obtain
\begin{eqnarray}
	\langle\sigma_1^{+}\sigma_1^{-}\rangle=\frac{NC\gamma+\Gamma}{2NC\gamma}.
\end{eqnarray}
These analytical expressions for the atom-atom correlation $\langle\sigma_{1}^{+}\sigma_{2}^{-}\rangle$ and the excited-state population $\langle\sigma_1^{+}\sigma_1^{-}\rangle$ are derived based on approximations that neglect non-collective terms, making them applicable in the enhanced radiance regime where $\gamma<\eta<NC\gamma$. Outside this  enhanced radiance regime, however, these expressions do not hold.

\section{Dicke-state population and atom-atom correlation in the low- and high-excitation regimes} 
\begin{figure}
	\centering
	\includegraphics[width=14cm]{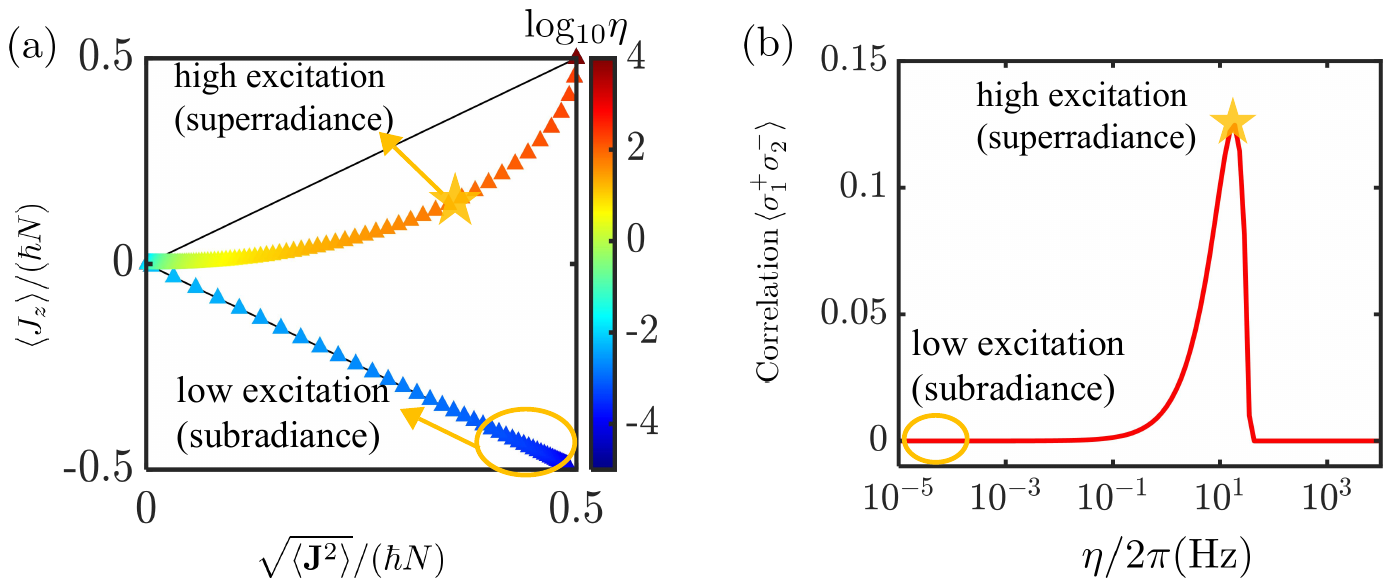}
	\caption{{(a) Dicke state population,  (b) atom-atom correlation $\langle\sigma_1^+\sigma_2^-\rangle$ in the steady-state versus $\eta$ at the EP. System parameters are $N={10}^7$, $\Delta_a=\Delta_b=0$, $g/2\pi=2.41\,\mathrm{Hz}$, $\kappa_a/2\pi=160\,\mathrm{kHz}$, $\kappa_b/2\pi=1\,\mathrm{kHz}$, $\gamma/2\pi=1\,\mathrm{mHz}$, and $\gamma_{\phi}/2\pi= 1\,\mathrm{mHz}$.}}\label{sv4}
\end{figure}

In this section, we present the Dicke-state population and atom-atom correlation in the low- and high-excitation regimes, which show that the atomic ensemble in our work cannot be simply described as only one atom effectively coupling to the cavity mode while all others remain in dark states. Specifically, the collective behavior of an ensemble of identical two-level systems can be conveniently described using the Dicke states $|J,M\rangle$. Figure\,\ref{sv4}(a) (corresponding to Fig.\,2(d) of the main text) shows the steady-state Dicke state population at the EP. As $\eta$ increases, the system transitions from a subradiance state (corresponding to the lower boundary of the Dicke ladder) to a superradiance state with low symmetry, and then ascends the Dicke ladder towards highly symmetric superradiance states, ultimately reaching the fully excited states. The lower-right corner of the Dicke triangle, also known as the low-excitation regime, is characterized by an average number of excited atoms $M+N/2$ much smaller than the total atom number $N$, which corresponds to $\langle J_z \rangle/(\hbar N) \approx -1/2$. The upper-right region of the Dicke triangle, corresponds to the high-excitation regime. Figure\,\ref{sv4}(b) (corresponding to Fig.\,2(b) of the main text) shows the variation of atom-atom correlations with increasing $\eta$. The atom-atom correlations first increase with $\eta$, reaching a maximum in the high-excitation regime, and then gradually decrease. 

\emph{I. Discussion of low-excitation regime.}---The lower-right corner of the Dicke triangle corresponds to the low-excitation regime, as shown in Fig.\,\ref{sv4}(a), where the incoherent pump is very weak (below $10^{-3}$). In the closed system, the atomic ensemble can be regarded as consisting of one bright mode and $N-1$ dark modes, with only the bright mode coupling to the cavity mode. However, in the open system considered here, the dephasing process intrinsically induces coupling between bright and dark modes, resulting in an indirect interaction between dark modes and the cavity mode\,\cite{SN. Lambert}.

We further clarify the physical mechanisms in both the closed and open systems. In the closed system, each two-level system can be approximated as a harmonic oscillator, valid when the excitation is weak\,\cite{ST. E. Li, SN. Moiseyev, SL. P. Lindoy}. The system Hamiltonian $H=\hbar \omega_a a^{\dagger} a+\hbar \omega_b b^{\dagger} b+\hbar G (a^{\dagger} b+a b^{\dagger})+\hbar \omega_\sigma \sum_{j=1}^N \sigma_j^{+} \sigma_j^{-}+\hbar g \sum_{j=1}^N\left(a^{\dagger} \sigma_j^{-}+a \sigma_j^{+}\right)$ can be written as $H=\hbar\omega_a a^{\dagger} a+\hbar \omega_b b^{\dagger} b+\hbar G(a^{\dagger} b+a b^{\dagger})+\hbar \omega_\sigma \sum_{j=1}^N c_j^{\dagger}c_j+\hbar g \sum_{j=1}^N\left(a^{\dagger} c_j+a c_j^{\dagger}\right)$, where $c_j$ and $c_j^\dag$ represent the bosonic annihilation and creation operators for atom $j$.  In the low excitation limit, we define the bright mode and $N-1$ dark mode operators as 
\begin{eqnarray}\label{4}
	B^{\dagger}=\frac{1}{\sqrt{N}} \sum_{j=1}^{N} c_{j}^{\dagger}, \qquad B=\frac{1}{\sqrt{N}} \sum_{j=1}^{N}c_j, \\\nonumber
	D_{k}^{\dagger}=\sum_{j=1}^{N} d_{kj}^* c_{j}^{\dagger}, \quad D_{k}=\sum_{j=1}^{N} d_{k j} c_j, \quad d_{k j}=\frac{1}{\sqrt{N}} \exp \left(\frac{2 \pi i k}{N} j\right).
\end{eqnarray}
The bright mode in the above expression corresponds to the special case $k=0$ of the dark-mode manifold. The bosonic approximation assumes that these modes obey bosonic commutation relations, that is 
\begin{eqnarray}
	[D_k,D_{k'}^{\dagger}]&=&[\sum_{j=1}^{N} d_{kj} c_{j},\sum_{j=1}^{N} d_{k'j}^* c_{j}^{\dagger}]\nonumber\\&=&\frac{1}{N}\sum_{j=1}^{N}\exp(\frac{2\pi ik}{N}j-\frac{2\pi ik'}{N}j)[c_j,c_j^{\dagger}]\nonumber\\&=&\delta_{k,k'}.
\end{eqnarray}
The transformed Hamiltonian takes the form
\begin{eqnarray}
	H=\hbar \omega_a a^{\dagger} a+\hbar \omega_b b^{\dagger} b+\hbar G (a^{\dagger} b+a b^{\dagger})+\hbar \omega_\sigma (B^{\dagger} B+\sum_{k=1}^{N-1} D_k^{\dagger} D_k)+\hbar g \sqrt{N}\left(a^{\dagger} B+a B^{\dagger}\right),
\end{eqnarray}
showing that only the bright mode couples to the cavity field. The bright mode represents a symmetric collective superposition of all atoms, not a single atom. In this case, the total emission rate scales linearly with $N$, as if each dipole emitted independently.

The above discussion applies to a closed quantum system in the low-excitation regime. In our work, we instead consider an open quantum system. To investigate the effects of dissipation, dephasing, and incoherent pumping on the dynamics of bright and dark modes within the bosonic approximation, we introduce the bright and dark mode excitation populations
\begin{eqnarray}
	n_b=\langle B^{\dagger} B \rangle, \qquad n_d=\langle \sum_{k=1}^{N-1} D_{k}^{\dagger} D_{k} \rangle.
\end{eqnarray}
Applying a rotating frame at frequency $\omega_{\sigma}$ and setting $\Delta_{a(b)}=\omega_{a(b)}-\omega_\sigma=0$, the Hamiltonian reduces to $H=\hbar G (a^{\dagger} b+a b^{\dagger})+\hbar g \sqrt{N}\left(a^{\dagger} B+a B^{\dagger}\right)$. Under the condition $\kappa_a,\kappa_b\gg\gamma$, the cavity mode can be adiabatically eliminated, yielding $b=-4gG\sqrt{N}/(\kappa_a\kappa_b)B$, $a=-2ig\sqrt{N}/\kappa_{a}B$. The dissipation terms of the two cavity modes in the master equation then simplify to $\kappa_a\mathcal{L}[a]\rho+\kappa_b\mathcal{L}[b]\rho=N\kappa_{ato}\mathcal{L}[B]\rho$, where $\kappa_{ato}=4g^2(4G^2+\kappa_a\kappa_b)/(\kappa_a^2\kappa_b)$. The Lindblad master equation thus becomes 
\begin{eqnarray}
	\frac{\partial}{\partial t}\rho&=&N\kappa_{ato}\mathcal{L}[B]\rho
	+\gamma\sum_{j=1}^{N}\mathcal{L}[\sigma_{j}^{-}]\rho	
	+\eta\sum_{j=1}^{N}\mathcal{L}[\sigma_{j}^{+}]\rho	
	+\gamma_{\phi}\sum_{j=1}^{N}\mathcal{L}[\sigma_{j}^{+} \sigma_{j}^{-}]\rho.\label{6}
\end{eqnarray}
In the low-excitation approximation, the atomic operator can be mapped to 
\begin{eqnarray}
	\sigma_{j}^{+}=c_j^{\dagger}= \sum_{k=0}^{N-1}d_{kj} D_{k}^{\dagger} \qquad \sigma_{j}^{-}=c_j= \sum_{k=0}^{N-1}d_{kj}^{*} D_{k},
\end{eqnarray}
where $k=0$ corresponds to the bright mode, and $k\neq 0$ correspond to the dark modes. The dissipative terms in the master equation can be expressed in terms of the bright and dark modes as 
\begin{eqnarray}
	\gamma\sum_{j=1}^{N}\mathcal{L}[\sigma_{j}^{-}]\rho\negthickspace&=&\negmedspace\negmedspace\gamma\sum_{j=1}^{N}\sum_{k=0}^{N-1}\sum_{k'=0}^{N-1}(d_{kj}^*D_{k}^{-}\rho d_{k'j}D_{k'}^{+}\negmedspace-\negmedspace\frac{1}{2}\rho d_{k'j}D_{k'}^{+}d_{kj}^*D_{k}^{-}\negmedspace-\negmedspace\frac{1}{2} d_{k'j}D_{k'}^{+}d_{kj}^*D_{k}^{-}\rho)
\end{eqnarray}
Due to 
\begin{eqnarray}
	\frac{1}{N}\sum_{j=1}^{N}\exp(\frac{2\pi ik'}{N}-\frac{2\pi ik}{N})j	=\delta_{k,k'},
\end{eqnarray}
we can obtain 
\begin{eqnarray}
	\gamma\sum_{j=1}^{N}\mathcal{L}[\sigma_{j}^{-}]\rho=\gamma\sum_{k=0}^{N-1}(D_{k}^{-}\rho D_{k}^{+}-\frac{1}{2}\rho D_{k}^{+}D_{k}^{-}-\frac{1}{2}D_{k}^{+}D_{k}^{-}\rho).\label{10}
\end{eqnarray}
Using the same approach, the incoherent pumping and dephasing terms can be expressed as
\begin{eqnarray}
	\eta\sum_{j=1}^{N}\mathcal{L}[\sigma_{j}^{+}]\rho \negthickspace&=&\negthickspace \eta\sum_{k=0}^{N-1}(D_{k}^{+}\rho D_{k}^{-}-\frac{1}{2}\rho D_{k}^{-}D_{k}^{+}-\frac{1}{2}D_{k}^{-}D_{k}^{+}\rho),\label{11}\\
	\gamma_{\phi}\sum_{j=1}^{N}\mathcal{L}[\sigma_{j}^{+}\sigma_{j}^{-}]\rho	\negthickspace&=&\negthickspace	\frac{\gamma_{\phi}}{N}\sum_{k=0}^{N-1}\sum_{k'=0}^{N-1}\sum_{m=0}^{N-1}(D_{k}^{+}D_{m}^{-}\rho D_{k'}^{+}D_{k+k'-m}^{-}\negthickspace-\negthickspace\frac{1}{2}\rho D_{k}^{+}D_{m}^{-}D_{k'}^{+}D_{k+k'-m}^{-}\label{12}\\&&-\frac{1}{2}D_{k}^{+}D_{m}^{-}D_{k'}^{+}D_{k+k'-m}^{-}\rho)\nonumber.
\end{eqnarray}
Substituting Eq.\,(\ref{10}), Eq.\,(\ref{11}), and Eq.\,(\ref{12}) into the master equation Eq.\,(\ref{6}), and taking $\langle o\rangle=tr(o\rho) $, we obtain the population dynamics of the bright and dark modes
\begin{eqnarray}
	\frac{dn_{b}}{dt}&=&-(N\kappa_{ato}+\gamma+\gamma_{\phi})n_{b}+\eta n_{b}+\eta,\\
	\frac{dn_{d}}{dt}&=&-\gamma n_{d}+\gamma_{\phi}n_{b}+\eta n_{d}+(N-1)\eta. \label{15}
\end{eqnarray}
The second term in the right of Eq.\,(\ref{15}) shows that dephasing mediates interconversion between the bright and dark modes. Consequently, the two modes become coupled in an open system. In principle, the bright mode couples directly to the cavity field, while the dark mode indirectly interacts with the cavity mode.

\emph{II. Discussion of high-excitation regime.}---The upper-right region of the Dicke triangle, as shown in Fig.\,\ref{sv4}(a), corresponds to the high-excitation regime. Our work focuses on this high-excitation regime, i.e, $\gamma<\eta<NC\gamma$, where atom–atom correlations increase and reach their maximum, yielding the minimal linewidth. In this high-excitation regime, the atomic ensemble is no longer confined to the lower-right corner of the Dicke triangle (low-excitation) but ascends the Dicke ladder towards the upper-right region of Dicke triangle (high-excitation). Consequently, the result Eq.\,(\ref{4}) based on the low-excitation approximation breaks down. As a result, the atomic ensemble can no longer be described in terms of uncoupled bright and dark modes. Instead, the atomic ensemble collectively couples to the cavity mode.

We further clarify the physical mechanisms in the high-excitation regime. In this regime, the atomic ensemble can not be treated as bosons, and we formally define the operators 
\begin{eqnarray}\label{17}
	B^{\dagger}=\frac{1}{\sqrt{N}} \sum_{j=1}^{N} \sigma_{j}^{\dagger}, \qquad B=\frac{1}{\sqrt{N}} \sum_{j=1}^{N}\sigma_j,\\\nonumber 
	D_{k}^{\dagger}=\sum_{j=1}^{N} d_{kj}^* \sigma_{j}^{\dagger}, \quad D_{k}=\sum_{j=1}^{N} d_{k j} \sigma_j, \quad d_{k j}=\frac{1}{\sqrt{N}} \exp \left(\frac{2 \pi i k}{N} j\right).
\end{eqnarray}
Correspondingly, the $B$ mode corresponds to the $k=0$ case of $D_k$ modes
\begin{eqnarray}
	[D_k,D_{k'}^{\dagger}]&=&[\sum_{j=1}^{N} d_{kj} \sigma_j^{-},\sum_{j=1}^{N} d_{k'j}^* \sigma_j^{+}]\nonumber\\
	&=&\frac{1}{N}\sum_{j=1}^{N}\exp(\frac{2\pi ik}{N}j-\frac{2\pi ik'}{N}j)[\sigma_j^{-},\sigma_j^{+}]\nonumber\\
	&=&-\frac{1}{N}\sum_{j=1}^{N}\exp(\frac{2\pi ik}{N}j-\frac{2\pi ik'}{N}j)\sigma_j^z.
\end{eqnarray}
Despite the commutation relations given above, the transformed Hamiltonian can still be written as 
\begin{eqnarray}
	H=\hbar \omega_a a^{\dagger} a+\hbar \omega_b b^{\dagger} b+\hbar G (a^{\dagger} b+a b^{\dagger})+\hbar \omega_\sigma (B^{\dagger} B+\sum_{k=1}^{N-1} D_k^{\dagger} D_k)+\hbar g \sqrt{N}\left(a^{\dagger} B+a B^{\dagger}\right).
\end{eqnarray}
To facilitate a qualitative discussion of the high-excitation regime, we neglect the fluctuation–dissipation terms. Based on the Heisenberg equation of motion, we obtain the dynamical evolution equation for the $B$ mode, i.e.,  $dB/dt=i/\hbar[H,B]+...=i\omega_{\sigma}B\sigma_j^z+ig\sqrt{N}a\sigma_j^z+i\omega_{\sigma}/N\sum_{k=1}^{N-1}\sum_{j=1}^{N}\allowbreak \exp(\frac{-2\pi ik}{N}j)\sigma_j^zD_k+...$.
The dynamical equation illustrates that the $B$ mode and $D_k$ modes interact with each other. Consequently, the $B$ and $D_k$ modes are coupled to the cavity mode, either directly or indirectly.

\section{Approximate analytical expression of spectrum linewidth obtained by the quantum regression theorem} \label{III}
In this section, we derive an analytical expression for the laser linewidth using the quantum regression theorem. The  linewidth $\Delta\nu$ can be obtained from the full width at half maximum\,(FWHM)\,of the laser power spectrum $S(\omega)$, which is related to the two-time correlation function $\langle a^{\dagger}(t)a(0)\rangle$ of the light field 
\begin{equation}
	S(\omega)=2\int_{0}^{\infty} \mathrm{Re}[\langle a^{\dagger}(t)a(0)\rangle e^{-i(\omega-\omega_{a})t}]\,dt.
\end{equation}
To obtain $\langle a^{\dagger}(t)a(0)\rangle$, we derive the following equations of motion using the quantum regression theorem
\begin{eqnarray}
	\frac{d}{dt}\langle a^{\dagger}(t)a(0)\rangle&=&(-\frac{1}{2}\kappa_{a}+i\Delta_{a})\langle a^{\dagger}(t)a(0)\rangle+iNg\langle\sigma_{1}^{+}(t)a(0)\rangle+iG\langle b^{\dagger}(t)a(0)\rangle,\\
	\frac{d}{dt}\langle
	\sigma_{1}^{+}(t)a(0)\rangle&=&-\frac{1}{2}(\gamma+\gamma_{\phi}+\eta)\langle\sigma_{1}^{+}(t)a(0)\rangle+ig\langle a^{\dagger}(t)a(0)\rangle-2ig\langle \sigma_{1}^{+}\sigma_{1}^{-}\rangle \langle a^{\dagger}(t)a(0)\rangle,\\
	\frac{d}{dt}\langle b^{\dagger}(t)a(0)\rangle&=&iG\langle a^{\dagger}(t)a(0)\rangle+(i\Delta_{b} -\frac{1}{2}\kappa_{b})\langle b^{\dagger}(t)a(0)\rangle.
\end{eqnarray}
Considering second-order approximation and factorizing  $\langle a^{\dagger}(t)a(0)\sigma_{1}^{+}(t)\sigma_{1}^{-}(t)\rangle\approx\langle\sigma_{1}^{+}(t)\sigma_{1}^{-}(t)\rangle \langle a^{\dagger}(t)a(0)\rangle$, we can obtain a closed set of equations of motion
\begin{equation}
	\frac{d}{dt}R(t)=\mathbb{T}\cdot R(t),\label{S20}
\end{equation}
where 
\begin{equation}
	R(t)=(\langle a^{\dagger}(t)a(0)\rangle,\langle\sigma_{1}^{+}(t)a(0)\rangle,\langle b^{\dagger}(t)a(0)\rangle)^{T},
\end{equation}
and 
\begin{eqnarray}
	\mathbb{T}=-\frac{1}{2}\begin{pmatrix}
		-\frac{1}{2}\kappa_{a}+i\Delta_{a}& \qquad  iNg &\qquad iG \\
		ig-2ig\langle\sigma_{1}^{+}\sigma_{1}^{-}\rangle &\qquad -\frac{1}{2}(\gamma+\gamma_{\phi}+\eta) &\qquad 0 \\
		iG                                  &\qquad 0 &\qquad i\Delta_{b}-\frac{1}{2}\kappa_{b}
	\end{pmatrix}.
\end{eqnarray}
The initial conditions for Eq. (\ref{S20}) are the steady-state solutions of the operators given in Section \ref{II}.

Since the matrix $\mathbb{T}$ is non-Hermitian, its eigenvalues are generally complex, and its eigenvectors are biorthogonal. To solve the dynamic equation Eq. (\ref{S20}), we introduce the left and right eigenvectors of $\mathbb{T}$ as 
\begin{eqnarray}
	\mathbb{T} |i\rangle =\lambda_{i}|i\rangle,\\
	\langle\tilde{i}|\mathbb{T} =\lambda_{i}\langle\tilde{i}|, 
\end{eqnarray}
where $\lambda_{i} (i=1,2,3)$ are the eigenvalues of $\mathbb{T}$, and $\langle \tilde{i}|$ and $|i\rangle$ are the left and right eigenvectors of $\mathbb{T}$, satisfying the biorthogonal relation $\langle \tilde{i}|j\rangle=\delta_{i,j}$ for $i,j=1,2,3$. The identity operator can then be expressed as $I=\sum_{i}|i\rangle\langle \tilde{i}|$.

Applying the Laplace transform to Eq. (\ref{S20}) gives
\begin{equation}
	\overline{R(p)}=\frac{1}{p-\mathbb{T}}R(0), \label{S25}
\end{equation}
where $\overline{R(p)}$ is the Laplace transform of $R(t)$.  By inserting the operator $I=\sum_{i}|i\rangle\langle \tilde{i}|$ into Eq. (\ref{S25}), we can obtain 
\begin{equation}
	\overline{R(p)}=\sum_{i}\frac{1}{p-\lambda_{i}}|i\rangle\langle\widetilde{i}|R(0).
\end{equation}
Performing the inverse Laplace transform on this equation, we find
\begin{equation}
	R(t)=\sum_{i}e^{\lambda_{i}t}|i\rangle\langle\widetilde{i}|R(0). \label{S27}
\end{equation}
From Eq. (\ref{S27}), we see that the two-time correlation function $\langle a^{\dagger}(t)a(0)\rangle$ is a superposition of the functions $e^{\lambda_{i}t}(i=1,2,3)$. After applying the Fourier transform to the correlation function, the laser spectrum can be derived as 
\begin{eqnarray} \label{S31-0}
	S(\omega)&=&2\sum_{i}\int_{0}^{\infty}dt\mathrm{Re}[C_{i}e^{\mathrm{Re}(\lambda_{i})t}e^{-i(\omega-\omega_{a}-\mathrm{Im}(\lambda_{i}))t}]\nonumber\\&=&\sum_{i} C_{i}\frac{2\mathrm{Re}(\lambda_{i})}{[\omega-(\omega_{a}+\mathrm{Im}(\lambda_{i}))]^{2}+(\mathrm{Re}(\lambda_{i}))^{2}}, 
\end{eqnarray}
where $S(\omega)$ is a superposition of three Lorentzian line shapes with central frequencies $\omega_{a}+\mathrm{Im}(\lambda_{i})$ and linewidths $2|\mathrm{Re}(\lambda_{i})|$ for $i=1,2,3$.

By solving the eigenfunction of $\mathbb{T}$ to first order, we obtain an approximate expression for the linewidth
\begin{eqnarray}
	\Delta\nu	=	2|\mathrm{Re}(\lambda_{i})|=\frac{(4G^{2}+\kappa_{a}\kappa_{b})(\gamma_{\phi}+\eta+\gamma)+4g^{2}N\kappa_{b}(1-2\langle\sigma_{1}^{+}\sigma_1^{-}\rangle)}{4G^{2}+(\kappa_{a}+\kappa_{b})(\gamma_{\phi}+\eta+\gamma)+\kappa_{a}\kappa_{b}+4g^{2}N(1-2\langle\sigma_{1}^{+}\sigma_1^{-}\rangle)}.
	\label{S29}
\end{eqnarray}
Substituting $\Gamma=\eta+\gamma_{\phi}+\gamma$, $\Gamma_c=4g^2\kappa_b/(4G^2+\kappa_a\kappa_b)$, and $\langle J_z\rangle/\hbar=N(\langle\sigma_1^{+}\sigma_1^{-}\rangle-1/2)$ into the Eq. (\ref{S29}), we can further simplify
\begin{eqnarray}
	\Delta\nu   &=& \frac{\Gamma+{4g^{2}N\kappa_{b}}(1-2\langle\sigma_{1}^{+}\sigma_1^{-}\rangle)/(4G^{2}+\kappa_{a}\kappa_{b})}{1+(\kappa_{a}+\kappa_{b})\Gamma/(4G^{2}+\kappa_{a}\kappa_{b})+4g^{2}N(1-2\langle\sigma_{1}^{+}\sigma_1^{-}\rangle)/(4G^{2}+\kappa_{a}\kappa_{b})}
	\nonumber\\	&=&	\frac{\Gamma+N\Gamma_{c}[1-2(\langle J_z\rangle/(\hbar N)+1/2)]}{1+(\kappa_{a}+\kappa_{b})\Gamma/(4G^{2}+\kappa_{a}\kappa_{b})+4g^{2}N[1-2(\langle J_z\rangle/(\hbar N)+1/2)]/(4G^{2}+\kappa_{a}\kappa_{b})}
	\nonumber\\&=&\frac{\Gamma-2\Gamma_{c}\langle J_z \rangle/\hbar}{1+(\kappa_{a}+\kappa_{b})\Gamma/(4G^{2}+\kappa_{a}\kappa_{b})-2\langle J_z \rangle\Gamma_{c}/(\hbar\kappa_{b})}.\label{S31}
\end{eqnarray}
The above Eq.\,(\ref{S31}) is Eq.\,(6) in the main text.
Substituting the EP condition $G=G_{\mathcal{PT}}=(\kappa_b-\kappa_a)/4$ into $\Gamma_c$ and $\Delta\nu$ in Eq. (\ref{S31}),  we obtain the linewidth at the EP 

\begin{eqnarray}
	\Delta\nu &=& 
	\frac{\Gamma-2\Gamma_c\langle J_z \rangle/\hbar}{1+4\Gamma/(\kappa_a+\kappa_b)-2\langle J_z \rangle \Gamma_c/(\hbar\kappa_b)},\label{47}
\end{eqnarray}
where $\Gamma_c=16g^2\kappa_b/(\kappa_a+\kappa_b)^2$. 

\begin{figure}
	\centering
	\includegraphics[width=13.5cm]{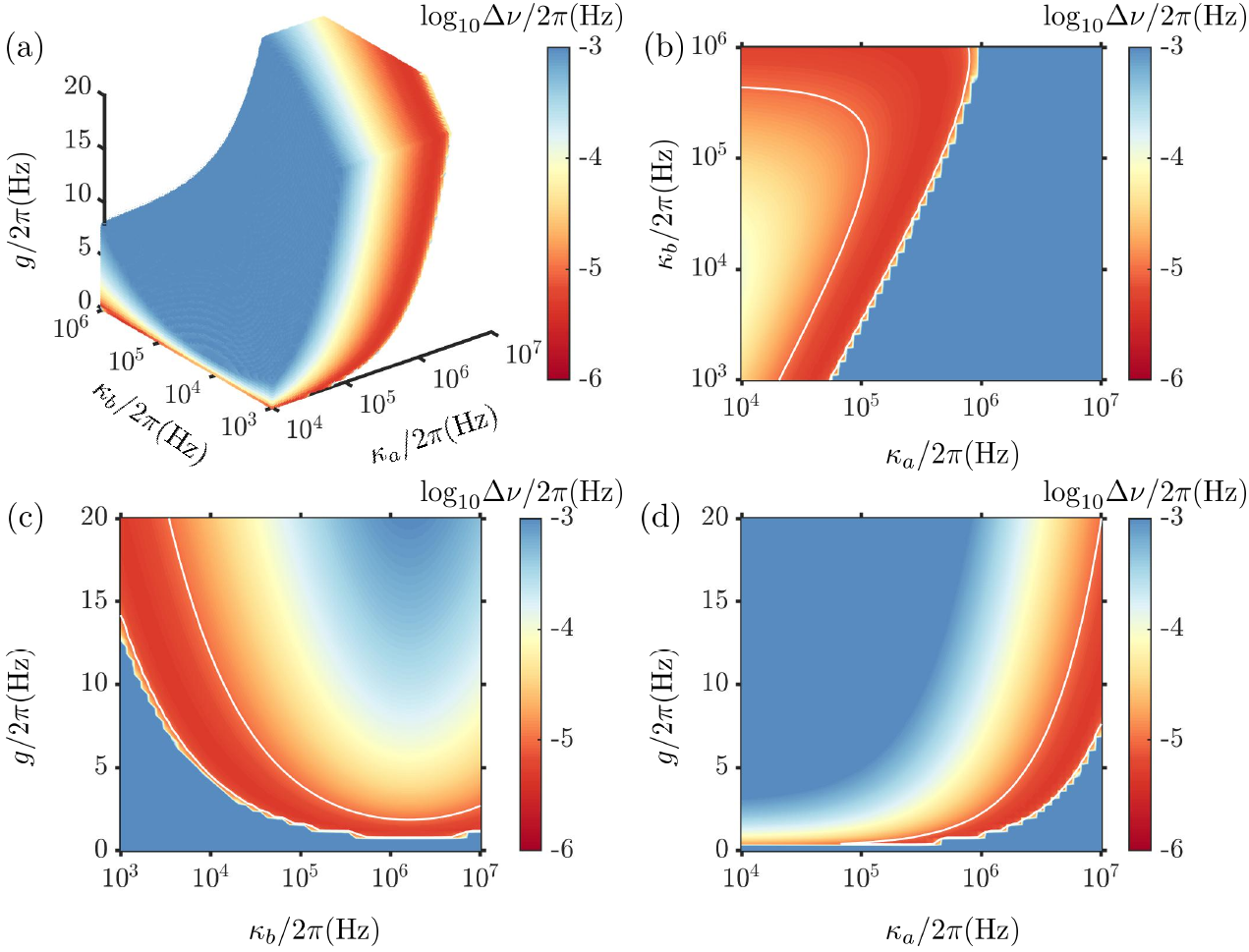}
	\caption{Emission linewidth $\Delta\nu$ at the EP versus cavity dissipations $\kappa_a$, $\kappa_b$, and the coupling strength $g$ for $\eta/2\pi=10\,\mathrm{Hz}$. For clearer visualization, panels (b)-(d) present special cuts along selected directions: (b) $g/2\pi=0.5\,\mathrm{Hz}$, (c) $\kappa_a/2\pi=1.59\,\mathrm{MHz}$, and (d) $\kappa_b/2\pi=0.14\,\mathrm{MHz}$. The white solid lines correspond to a linewidth of $10^{-5}\,\mathrm{Hz}$.}\label{sv3}
\end{figure}

Equation\,(\ref{47}) shows that, in addition to the intrinsic system parameters, the linewidth at the EP is affected by $\Gamma$ and $\langle J_z\rangle$. The formula is highly complex, making it difficult to analytically determine the fundamental limit. Therefore, within experimentally feasible parameter ranges, we numerically discuss the minimum value of linewidth $\Delta\nu$ by investigating the effects of the dissipation rates $\kappa_a$,  $\kappa_b$ and the atom–cavity coupling strength $g$. As shown in Fig.\,3(b) of the main text, the linewidth at the EP decreases and then increases with increasing the incoherent pumping. This behavior arises from the corresponding growth and reduction of atom–atom correlations, shown in Fig.\,2(b) of the main text. In our analysis, we set $\eta=10\,\mathrm{Hz}$, close to the value that maximizes atomic coherence and minimizes the linewidth.

Under the condition of $\eta=10\,\mathrm{Hz}$, Figure\,\ref{sv3}(a) depicts the linewidth at the EP as a function of $\kappa_a$, $\kappa_b$, and $g$. The results reveal that, by appropriately selecting the coupling and dissipation strengths at the EP, the linewidth can be maintained at approximately $2\pi \times 10^{-6}\,\mathrm{Hz}$. To visualize it, we extract from Fig.\,\ref{sv3}(a) the parameter points that both yield the minimal linewidth and satisfy  the bad-cavity condition $\kappa_a, \kappa_b \gg g\sqrt{N}$ ($g/2\pi=0.5\,\mathrm{Hz}$, $\kappa_a/2\pi=1.59\,\mathrm{MHz}$, $\kappa_b/2\pi=0.14\,\mathrm{MHz}$). Figures\,\ref{sv3}(b)-\ref{sv3}(d) show two-dimensional cross sections around three points. Figure\,\ref{sv3}(b) shows that although the bad-cavity regime requires large cavity dissipation, increasing $\kappa_{a,b}$ indefinitely broadens the linewidth. This is because the cooperativity $C=4g^2/(\kappa_{\mathrm{eff}}\gamma)$, which quantifies the relative strength of the atom–cavity coupling compared with dissipative processes, decreases as the cavity dissipation increases. When $C$ becomes too small, atom-atom correlations diminish, causing the linewidth to broaden. Figures\,\ref{sv3}(c) and \,\ref{sv3}(d) show that the coupling strength $g$ must lie in an appropriate range. An excessively large $g$ violates the bad-cavity condition $g\sqrt{N} \ll \kappa_a, \kappa_b$, while excessively small $g$ lowers the cooperativity $C$ and reduces atom–atom correlations, again increasing the linewidth. Therefore, by choosing appropriately system parameters $\kappa_a$, $\kappa_b$ and $g$, the minimum linewidth at the EP can be maintained on the order of $10^{-6}\,\mathrm{Hz}$ (corresponding to the areas enclosed by the white curves).

\section{Calculation of spectrum with a filter cavity}
To obtain the laser spectrum, we can utilize two methods. The first involves using the quantum regression theorem combined with a Fourier transform to derive an analytical lasing spectrum. The second method utilizes a filter cavity, where the cavity output is coupled to the filter cavity, and the excitation spectrum of the filter cavity is used to characterize the output spectrum of the main cavity, i.e., the lasing spectrum. We modify the master equation in the main text by adding the terms $-i/ \hbar[H_{f}+H_{f-c},\rho]+\kappa_{f}\mathcal{L}[f]\rho$\,\cite{SY. Zhang}. Photon loss within the filter cavity is characterized by a Lindblad term with decay rate $\kappa_{f}$. Here, the filter cavity Hamiltonian is given by $H_{f}=\omega_{f}f^{\dagger}f$, where $\omega_{f}$ is the frequency of the filter cavity, and $f^{\dagger}$ and $f$ are the creation and annihilation operators of the cavity. The interaction between the filter cavity and the main system  is described by  $H_{f-c}=\beta(f^{\dagger}a+a^{\dagger}f)$, with coupling strength $\beta$. 

\begin{figure}
	\centering
	\includegraphics[width=16cm]{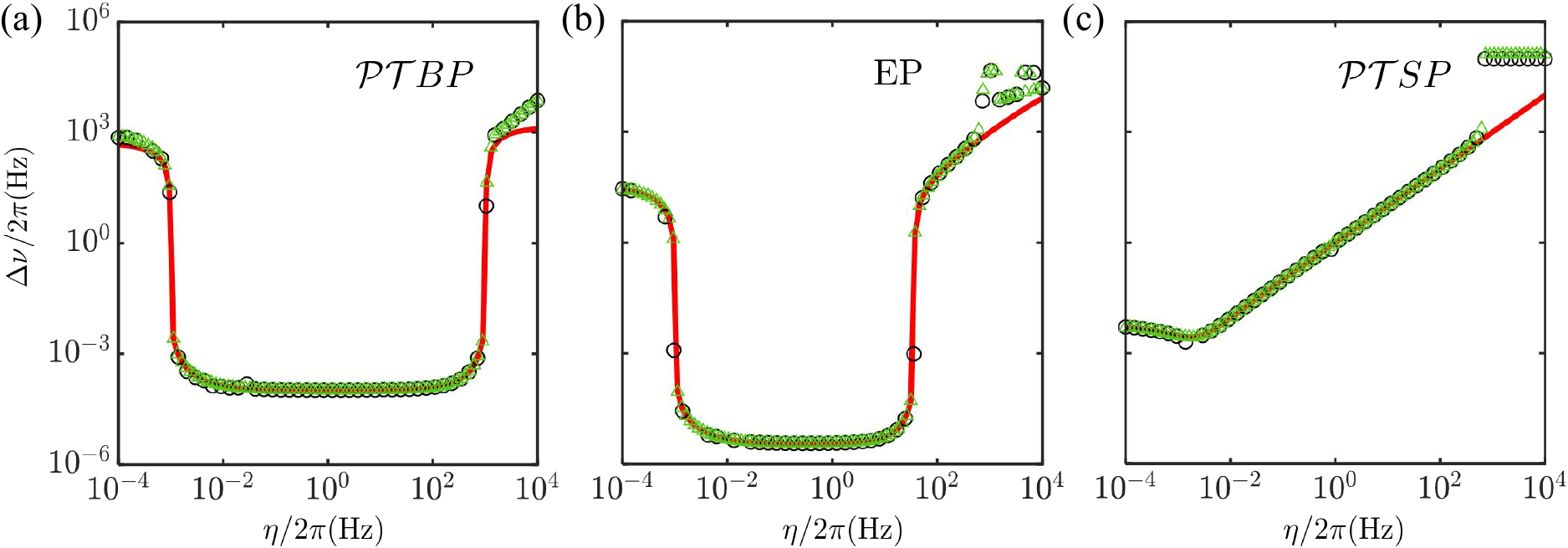}
	\caption{Laser linewidth $\Delta\nu$ vs $\eta$ in the (a) $\mathcal{PT}$BP, (b) EP, and (c) $\mathcal{PT}$SP. Green triangles and red solid lines correspond to results obtained by the derivation from Eq.(\ref{S31-0}) and the analytical expression Eq.(\ref{S31}) based on the quantum regression theorem, respectively. Black circles represent the results obtained using the filter cavity method. The system parameters are the same as in Fig.\,3(b).}\label{sv1}
\end{figure}

To obtain the spectrum, we calculate the steady-state photon number $\langle f^{\dagger} f \rangle$ in the filter cavity for various values of $\omega_f$, using second-order mean-field theory\,\cite{SY. Zhang}. By fitting the spectrum with a Lorentzian function, we extract the linewidth $\Delta \nu$. The mean photon number in the filter cavity evolves according to 
\begin{eqnarray}
	\frac{d}{dt}\langle f^{\dagger} f\rangle&=&-2\beta\mathrm{Im}\langle a^{\dagger}f\rangle-\kappa_{f}\langle f^{\dagger}f \rangle,
\end{eqnarray}
which includes the correlation $\langle a^{\dagger}f\rangle$. The evolution equation for this correlation is 
\begin{eqnarray}
	\frac{d}{dt}\langle a^{\dagger}f\rangle&=&iG\langle b^{\dagger}f\rangle-\frac{1}{2}(\kappa_{a}+\kappa_{f})\langle a^{\dagger}f\rangle+(i\Delta_{a}-i\delta)\langle a^{\dagger}f\rangle+i\beta(\langle f^{\dagger}f\rangle-\langle a^{\dagger}a\rangle)+iNg\langle\sigma_{1}^{+}f\rangle,
\end{eqnarray}
where $\delta=\omega_f-\omega_{\sigma}$ denotes the frequency detuning of the filter cavity mode $f$ and the atoms. This equation depends on the correlations $\langle \sigma_{1}^{+}f\rangle$ and $\langle b^{\dagger}f\rangle$, governed by
\begin{eqnarray}
	\frac{d}{dt}\langle
	b^{\dagger}f\rangle&=&-\frac{1}{2}(\kappa_{b}+\kappa_{f})\langle b^{\dagger}f\rangle+(i\Delta_{b}-i\delta)\langle b^{\dagger}f\rangle+iG\langle a^{\dagger}f\rangle-i\beta\langle ab^{\dagger}\rangle,\\	\frac{d}{dt}\langle\sigma_{1}^{+}f\rangle&=&ig\langle a^{\dagger}f\rangle-i\beta\langle a\sigma_{1}^{+}\rangle-\frac{1}{2}(\Gamma+\kappa_{f})\langle\sigma_{1}^{+}f\rangle-i\delta\langle\sigma_{1}^{+}f\rangle-2ig\langle\sigma_{1}^{+}\sigma_{1}^{-}\rangle\langle a^{\dagger}f\rangle. 	
\end{eqnarray}
By combining the above four equations with the equations of motion for the operator averages given in Section \ref{II}, a closed set of equations can be obtained, allowing the steady-state photon number $\langle f^{\dagger}f\rangle$ in the filter cavity to be determined, and subsequently, the spectrum can be derived. Additionally, the inclusion of the filter cavity terms in the master equation causes changes in some of the equations in Section \ref{II}. The equation for  intracavity photon number becomes
\begin{eqnarray}
	\frac{d}{dt}\langle a^{\dagger}a\rangle&=&2\beta \mathrm{Im}\langle a^{\dagger}f\rangle+2G\mathrm{Im}\langle a^{\dagger}b\rangle+2gN\mathrm{Im}\langle a^{\dagger}\sigma_{1}^{-}\rangle-\kappa_{a}\langle a^{\dagger}a\rangle.
\end{eqnarray}
The correlation between the two cavity modes, as well as the correlation between the atom and the cavity mode, are updated to
\begin{eqnarray}
	\frac{d}{dt}\langle
	a^{\dagger}b\rangle&=&i\beta\langle bf^{\dagger}\rangle-\frac{1}{2}(\kappa_{a}+\kappa_{b})\langle a^{\dagger}b\rangle+iG(\langle b^{\dagger}b\rangle-\langle a^{\dagger}a\rangle)+i(\Delta_{a}-\Delta_{b})\langle a^{\dagger}b\rangle+iNg\langle b\sigma_{1}^{+}\rangle,\\
	\frac{d}{dt}\langle a^{\dagger}\sigma_{1}^{-}\rangle&=&i\beta\langle f^{\dagger}\sigma_{1}^{-}\rangle-\frac{1}{2}(\Gamma +\kappa_{a})\langle a^{\dagger}\sigma_{1}^{-}\rangle+ig\langle\sigma_{1}^{+}\sigma_{1}^{-}\rangle+iG\langle b^{\dagger}\sigma_{1}^{-}\rangle-ig\langle a^{\dagger}a\rangle\nonumber\\&&+i\Delta_{a}\langle a^{\dagger}\sigma_{1}^{-}\rangle+ig(N-1)\langle\sigma_{1}^{+}\sigma_{2}^{-}\rangle+2ig\langle\sigma_{1}^{+}\sigma_{1}^{-}\rangle\langle a^{\dagger}a\rangle.
\end{eqnarray}
To minimize the impact of the filter cavity on the main system, we assume a very small value for $\beta$. Furthermore, to accurately resolve the spectrum, we choose $\kappa_f$ to be much smaller than the linewidth of the steady-state spectrum. Figure\,\ref{sv1} compares the emission linewidths obtained using two methods: one based on the quantum regression theorem combined with a Fourier transform and the other employing a filter cavity to analyze the cavity output. The results from both approaches are consistent, even when compared with the analytical emission linewidths.

\begin{figure}
	\centering
	\includegraphics[width=17cm]{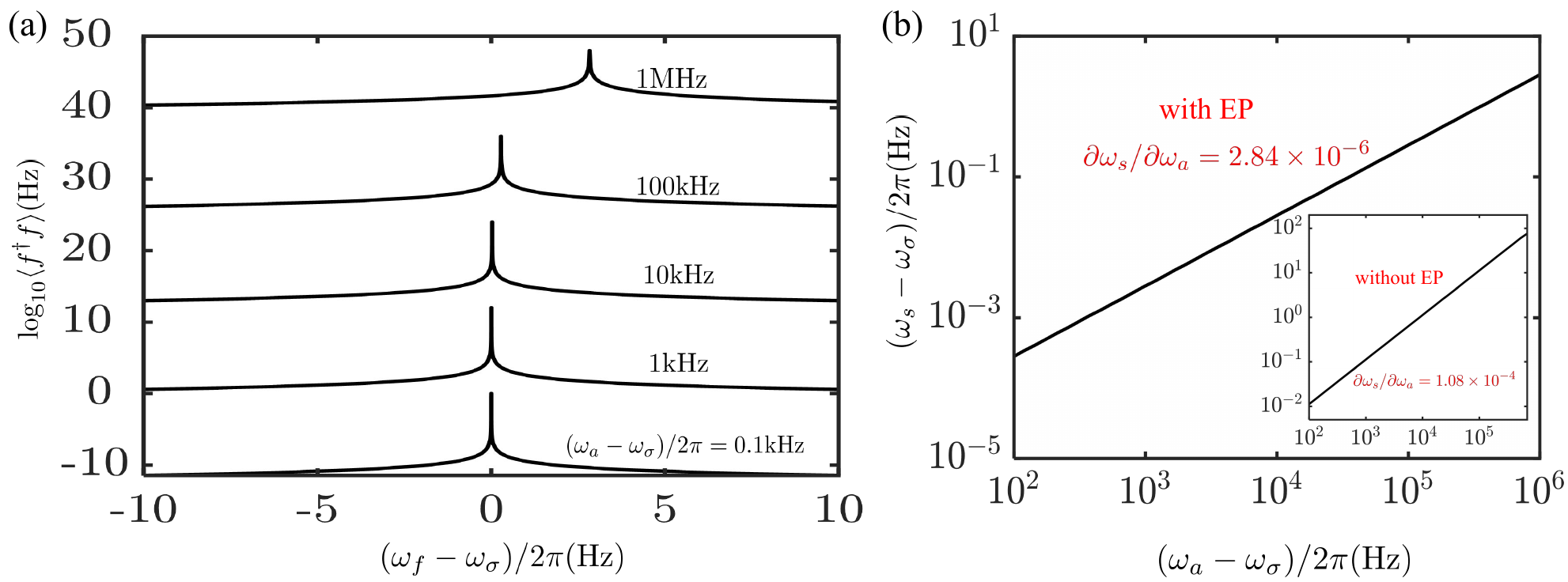}
	\caption{Cavity pulling in superradiant lasing. Panel (a) shows that the variation in normalized steady-state spectra,  presented on a logarithmic scale,  as a function of the detuning between the filter cavity frequency and the atomic frequency in the EP regime, with $G/2\pi=39.75\,\rm{kHz} $. For clarity, spectra are vertically shifted for different detuning values of the cavity mode frequency with respect to the bare atomic frequency, specified as  $\omega_{a\sigma}=(\omega_a-\omega_{\sigma})/2\pi=0.1,1,10,100,1000$\,kHz. This analysis assumes an  incoherent pump rate $\eta/2\pi=18\,\rm{Hz}$ and an atom number $N=10^7$. Panel (b) shows the peak frequency of the lasing spectrum as a function of detuning of $\omega_{a\sigma}$ for systems with EP and without EP (inset; $\kappa_b=0$ and $G=0$).} \label{sv}
\end{figure}

\section{Frequency pulling by cavity detuning and atom number fluctuations by technical noises}\label{VI}
It is common that the frequency of the cavity mode frequency fluctuates in experiments, due to factors such as vibrations of mirrors constituting the cavity, which in turn causes variations in the laser output frequency. However, superradiant lasing primarily operates in the bad-cavity regime, where the cavity dissipation rate is significantly higher than the atomic dissipation rate. In this regime, the central frequency of the output lasing is determined by the atomic transition frequency rather than the cavity’s central frequency. This dependency makes the lasing frequency robust against fluctuations in the cavity resonance frequency. These fluctuations are transferred to the laser output frequency, which can be characterized by a factor known as the cavity pulling factor $P$, defined as $P=\partial\omega_s/\partial\omega_{a}$.
This factor measures the change in the laser frequency relative to the cavity frequency.

In this section, we evaluate the cavity pulling factor $\partial\omega_s/\partial\omega_{a}$ using the method of coupling the cavity output to a filter cavity, which quantifies the laser
frequency shift $\partial\omega_s$ in response to small fluctuations $\partial\omega_{a}$ in the cavity mode frequency\,\cite{SM. A. Norcia, SJ. G. Bohnet, SH. Liu, SM. A. Norcia1}. To investigate this, we compute the normalized lasing spectrum $\langle f^{\dagger}f\rangle$ on a logarithmic scale as a function of detuning $\omega_{a\sigma}=(\omega_a-\omega_{\sigma})/2\pi$ in Fig.\,\ref{sv}(a), using values from $0.1\,{\rm kHz}$ to $1000\,{\rm kHz}$ between the cavity field and atoms. For this analysis, we consider a system with $N=10^7$ atoms, an incoherent pumping rate $\eta/2\pi=18$\,Hz, and a tunneling coupling $G/2\pi=39.75\,\rm{kHz}$ (corresponding to EP). Across all detuning values, the lasing spectrum exhibits a dominant peak at the filter cavity resonance frequency $\omega_f$, with these peaks shifting to higher frequencies as detuning increases. The peak frequency of the lasing spectrum as a function of detuning of $\omega_{a\sigma}$ can be extracted from the data shown in Fig.\,\ref{sv}(a). We also compared the variations of the output laser’s peak frequency induced by the cavity pulling effect in systems without and with an EP. Figure\,\ref{sv}(b) and its inset show the peak frequency of the laser spectrum as functions of the cavity resonance frequency for both cases. We find that, in either case, the peak frequency changes approximately linearly and monotonically with the cavity resonance frequency. Specifically, in the presence of the EP, the frequency sensitivity $\partial\omega_s/\partial\omega_a$ is $2.84\times10^{-6}$, whereas in the absence of the EP it is $1.08\times10^{-4}$. These results demonstrate that our EP-based scheme exhibits two orders of magnitude greater robustness against external cavity noise, which is highly advantageous for improving the frequency stability of active optical clocks.

\begin{figure}
	\centering
	\includegraphics[width=8.0cm]{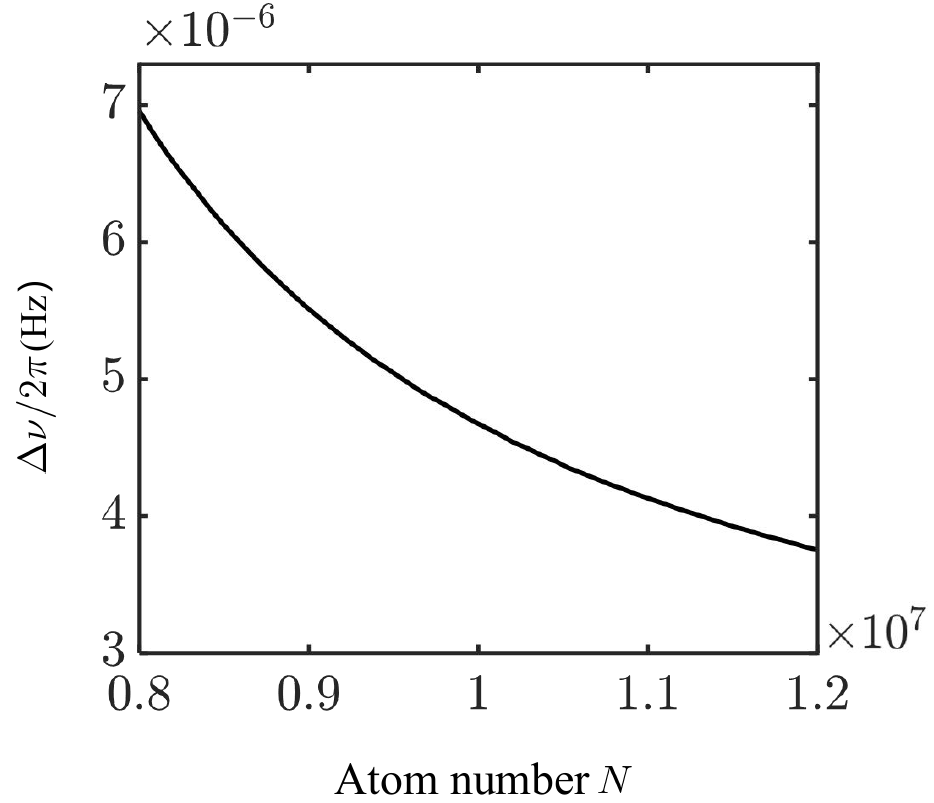}
	\caption{Emission linewidth $\Delta\nu$ versus the atom number $N$ in the EP, when $G/2\pi=39.75\mathrm{kHz}$ and $\eta/2\pi=18\,\mathrm{Hz}$.}\label{sv2}
\end{figure}

In this section, we also evaluate the effect of atom-number fluctuations on the final linewidth of superradiant lasing by coupling the cavity output to a filter cavity, which enables us to quantify the linewidth broadening induced by fluctuation in $N$. For an ensemble of approximately $N={10}^7$ atoms, the relative atom number fluctuations are typically on the order of $20\%$\,\cite{SM. A. Norcia}. Figure\,\ref{sv2} shows the dependence of the linewidth $\Delta\nu$ on the atom number around $N={10}^7$, assuming incoherent pumping $\eta/2\pi=18\mathrm{Hz}$ and tunneling coupling $G/2\pi=39.75\,\mathrm{kHz}$. It is shown that the linewidth decreases weakly and monotonically with increasing $N$. Within the experimentally relevant range $N=\left(1.0\pm0.2\right)\times{10}^7$, the linewidth variation is approximately $3.2\times{10}^{-6}\,\mathrm{Hz}$ at the EP. This demonstrates that, in order to achieve a linewidth approaching the  $\mu\mathrm{Hz}$ level at the EP, atom-number fluctuations should be kept below 20$\%$, which is experimentally feasible.

\section{Experimental feasibility and stability analysis of different optical atomic clocks.}
In this section, we provide a detailed analysis explaining why our work is realistic enough to motivate experimentalists to pursue the development of a new type of optical clock that would be more stable than current ones. 

\emph{I. Discussion of experimental feasibility.}---In principle, our model is experimentally feasible based on recent experimental progress. On one hand, pulsed superradiant lasing has already been realized on the highly forbidden millihertz-linewidth clock transition $^{3}P_0$ to $^{1}S_0$ in an ensemble of laser-cooled $^{87}\mathrm{Sr}$ atoms trapped within a high-finesse optical cavity\,\cite{SJ. G. Bohnet, SM. A. Norcia1, SM. A. Norcia,ST. Laske, SS. Dubey, SM. A. Norcia2}. Specifically, up to $N=2.5\times10^5$ $^{87}\mathrm{Sr}$ atoms are cooled to $10\,\mathrm{\mu K}$ and tightly trapped along the axis of a high-finesse optical cavity  ($F=2.4\times10^4$; linewidth, $\kappa/2\pi=160\,\mathrm{kHz}$)  by an optical lattice near the magic wavelength of $813.4274\,\mathrm{nm}$, where the two clock states experience identical frequency shifts, making the transition frequency independent of lattice intensity\,\cite{SH. Katori}. 
The realization of superradiant laser primarily involves the following steps. Atoms are first prepared via two-stage cooling: initial capture and cooling using the dipole-allowed $^{1}S_0$ to $^{1}P_1$ transition at $461\,\mathrm{nm}$, followed by final cooling and lattice loading via the narrow $^{1}S_0$ $F=11/2$ to $^{3}P_1$ $F=11/2$ transition at $689\,\mathrm{nm}$. Once loaded into the lattice, a magnetic field of several Gauss perpendicular to the cavity defines the quantization axis. The atoms are then optically pumped to the $m_f=9/2$ state using a circularly polarized pump beam near the resonance with the $^{1}S_0$  to $^{3}P_1$, $F=9/2$ transition. Subsequently, a $698\,\mathrm{nm}$ laser sideband along the cavity axis adiabatically transfers the atoms to the $^{3}P_0$ state. To detect the superradiant laser pulses, the cavity output is coupled to a single-mode fiber and detected using a single-photon counting module (SPCM) whose transistor-transistor logic (TTL) output is low-pass filtered to provide a signal proportional to the photon emission rate.  After a fixed duration (typically $T=300\,\mathrm{ms}$), the atomic population is measured via a resonant $461\,\mathrm{nm}$ fluorescence beam and a charge-coupled device (CCD) camera. To investigate the spectral properties of the superradiant laser pulses, a beat note can be formed between two nearby Zeeman transitions of the strontium atoms. Specifically, the optical-pumping efficiency is intentionally reduced so that atoms populate both the $m_f=9/2$ and $m_f=7/2$ ground states. The atoms are then adiabatically transferred into ground–excited superposition states with different $m_f$ projections. This produces two independent atomic subensembles that interact with the same cavity mode but possess slightly different transition frequencies. As a result, a modulation appears in the emitted power at a frequency equal to the magnetic-field-induced splitting between the two transitions. This amplitude modulation arises from beating between the adjacent Zeeman transitions. By performing a Fourier transform of the emitted power and fitting the spectral peak with a Lorentzian profile, the full width at half maximum (FWHM) of the beat note can be extracted. The above procedure can be used to realize the pulsed superradiant lasing. However, realizing continuous superradiant lasing remains challenging, since it is experimentally difficult to continuously load a high flux of cold atoms into a ring cavity. A recent breakthrough demonstrated continuous loading of $2.1\times10^7$ $^{88}\mathrm{Sr}$ atoms/s into a ring cavity in the strong collective coupling regime, and subsequent transport of the atoms in a moving intracavity lattice to a low-decoherence region, which paves the way for the realization of a continuous superradiant laser on the optical clock transition in strontium\,\cite{SJ. R. K. Cline1}.

On the other hand, the exceptional points (EPs) have been experimentally realized using two coupled Fabry–P\'{e}rot (FP) microcavities\,\cite{SM. Fan,SZ. Li}. Specifically, the coupled FP microcavities consist of identical fiber Bragg gratings (FBGs), with $Er^{+}_{3}$- and $Ce^{3+}$-doped phosphosilicate sol–gel coatings serving as gain and loss media, respectively. Owing to strong reflections at the interfaces between the gain, coupling, and loss regions, two optical cavities are formed within the system. The coupling strength between the two microcavities depends on both their separation distance and the refractive index (RI) of the coupling region. The entire structure is encapsulated within a silica capillary with T-shaped junctions serving as inlets and outlets for functional liquids,  constructing a microfluidic system. By tuning the RI of the coupling region, one can precisely control the system’s phase and the EP position. In our model, the EP arises from  two dissipatively coupled FP cavities. Hence, each FBG facet can be coated with a lossy material (e.g., $Ce^{3+}$-doped phosphosilicate sol–gel) to realize this configuration. Consequently, combining the above two types of experiments—superradiant lasing in strontium and EP formation in coupled FP cavities—makes our proposed scheme experimentally feasible.

\emph{II. Discussion of stability of different optical atomic clocks.}---Our model can be utilized to realize superradiant lasing with ultranarrow linewidth. Such superradiant lasing holds promising applications in optical atomic clocks that would be more stable than current ones. In principle, the stability of an optical atomic clock is limited by the standard quantum limit (SQL),  which arises from quantum projection noise (QPN)\,\cite{SW. M. Itano}. The QPN-limited fractional frequency instability can be expressed as\,\cite{ST. L. Nicholson}
\begin{eqnarray}\label{54}
	\sigma_{\mathrm{QPN}}(\tau)=\frac{\chi\Delta\nu}{\pi \omega_\sigma} \sqrt{\frac{T_c}{N\tau}},
\end{eqnarray}
where $\sigma_{\rm QPN}(\tau)$ is the QPN-limited fractional instability of a clock, $\omega_\sigma$ is the clock transition frequency, $\Delta\nu$ is the excitation linewidth, $N$ is the number of atoms, $T_c$ is the clock cycle time, $\tau$ is the averaging time, and $\chi$ is a numerical factor near unity that is determined by the line shape of the clock transition spectroscopy. In practice, optical lattice clocks typically operate with an instability above this limit due to noise from the probe laser used to interrogate the atomic transition. A commonly experimental measure of clock noise performance is the fractional Allan deviation\,\cite{SLudlow2015BY, SE. Oelker, SM. Takamoto}
\begin{eqnarray}
	\sigma(\tau)=\sqrt{\frac{1}{(N-1)} \sum_{n=1}^{N-1} \frac{\left(\bar{\omega}_{n+1}-\bar{\omega}_n\right)^2}{2 \omega_\sigma^2}},
\end{eqnarray}
where $\tau=T_c$ corresponds to a single experimental cycle, $\bar{\omega}_n$ denotes the frequency estimate obtained in the $n\mathrm{th}$ measurement, and $\omega_\sigma$ is the reference atomic transition frequency. This expression quantifies the frequency instability by evaluating the variance of frequency differences between consecutive measurements.

Recent synchronous clock comparisons suppress local oscillator (Dick) noise, achieving a fractional frequency instability of $10^{-18}$ at $\tau=1\,\mathrm{s}$\,\cite{ST. Bothwell}. Active optical clocks based on superradiant lasers are still at the proof-of-principle stage, as continuous superradiant operation has not yet been realized. Nonetheless, a ring cavity that continuously loads atoms and transports them in a light conveyor belt to a low-decoherence region has been realized\,\cite{SJ. R. K. Cline1}. This marks a crucial step toward continuous superradiant lasing and more stable active optical clocks. Based on numerical results from  seminal theoretical studies of superradiant lasers\,\cite{SD. Meiser}, the fractional frequency instability (determined by the SQL) of an active optical clock can be estimated. For representative parameters of  $\omega_\sigma/2\pi\approx 10^{14}\,\mathrm{Hz}$, $N\approx 10^6$, $\Delta\nu/2\pi\approx 10^{-3}\,\mathrm{Hz}$, $\chi\approx1$, and $T_c\approx 1\,\mathrm{s}$, the theoretical limit is approximately $10^{-21}$ at $\tau=1\,\mathrm{s}$, which is three orders of magnitude smaller than the instability currently achieved by passive optical clocks $10^{-18}$ at $\tau=1\,\mathrm{s}$\,\cite{ST. Bothwell}.

Here, we propose to achieve superradiant lasing with an ultranarrow linewidth in a PT-symmetric system. The linewidth of superradiant lasing can reach the order of $\mu\,\mathrm{Hz}$ at the EP. Based on the SQL, i.e., Eq.\,(\ref{54}), the instability of the corresponding active optical clock is estimated to be $10^{-24}$ at $\tau=1\,\mathrm{s}$ for $\omega_\sigma/2\pi \approx 10^{14}\,\mathrm{Hz}$, $N\approx 10^7$, $\Delta\nu/2\pi\approx 10^{-6}\,\mathrm{Hz}$, $\chi\approx 1$, and $T_c\approx 1\,\mathrm{s}$. This value is three orders of magnitude lower than the instability $10^{-21}$ at $\tau=1\,\mathrm{s}$ estimated in the absence of the EP, which reflects the significant improvement in clock stability resulting from linewidth suppression at the EP. In practice, the fractional frequency instability is slightly larger than the SQL due to technical noise sources such as cavity mirror vibrations, which can be characterized by the cavity pulling effect.  
In Section \ref{VI}, we show that the cavity pulling factor at the EP is $2.84\times{10}^{-6}$, which is smaller than the  corresponding value $1.08\times{10}^{-4}$ in the absence of the EP. These results demonstrate that our scheme can realize optical atomic clocks with lower frequency instability and enhanced robustness against external noise.

\end{document}